\documentclass[lettersize,journal]{IEEEtran}

\usepackage[margin=1.in]{geometry}
\usepackage{authblk}  

\usepackage{amsmath}

\usepackage{amsthm}
\usepackage{amssymb}
\usepackage{amsfonts}
\usepackage{dsfont}
\usepackage{accents}
\usepackage{bm}
\usepackage{hyperref}

\usepackage{tabularx} 
\usepackage{caption}
\usepackage{multirow}
\usepackage{booktabs}
\usepackage{graphicx}
\graphicspath{{Figures/}}
 \usepackage{subcaption}
 \usepackage{tcolorbox}
 \usepackage{wasysym}
  \usepackage{longtable}

\usepackage{tikz,pgfplots}

\usepgfplotslibrary{groupplots}
\usetikzlibrary{intersections,calc,arrows,matrix,spy}
\usepackage{color}
\definecolor{darkred}{rgb}{0.6,0,0}
\definecolor{darkgreen}{rgb}{0,0.5,0}
\definecolor{darkblue}{rgb}{0,0,0.5}
\definecolor{SkyBlue}{rgb}{0.53, 0.81, 0.92}
\pgfplotsset{compat=1.5.1}

\usepackage{algorithm}
\usepackage{algpseudocode}
\usepackage[squaren,Gray]{SIunits}

\usepackage[normalem]{ulem}
\usepackage{cleveref}

\definecolor{mycolor1}{RGB}{17, 17, 246}
\definecolor{mycolor2}{RGB}{158, 8, 219}
\definecolor{mycolor3}{RGB}{207,9,9}
\definecolor{mycolor4}{RGB}{28,190,6} 
\definecolor{mycolor5}{RGB}{17, 17, 246}
\definecolor{mycolor6}{RGB}{100,0,170}
\definecolor{mycolor7}{RGB}{64, 224, 208}%
\definecolor{mycolor8}{RGB}{50, 205, 50}%
\definecolor{mycolor9}{RGB}{139, 0, 0}%
\definecolor{mycolor10}{RGB}{255, 215, 0}%

\def\method{ICE-TIDE}

\def\R{\mathbb{R}}          									 	          

\newcommand{\xb}{\mathrm{\mathbf{x}}}
\newcommand{\yb}{\mathrm{\mathbf{y}}}

\newcommand{\Cb}{\mathrm{\mathbf{C}}}

\newcommand{\Pb}{\mathrm{\mathbf{P}}}

\newcommand{\Rb}{\mathrm{\mathbf{R}}}

\newcommand{\Vb}{\mathrm{\mathbf{V}}}

\newcommand{\gammab}{\mathrm{\bm{\gamma}}}

\newcommand{\etab}{\mathrm{\bm{\eta}}}

\newcommand{\rhob}{\mathrm{\bm{\rho}}}

\newcommand{\taub}{\mathrm{\bm{\tau}}}

\newcommand{\psib}{\mathrm{\bm{\psi}}}

\newcommand{\xib}{\mathrm{\bm{\xi}}}

\newcommand{\Phib}{\mathrm{\bm{\Phi}}}

\newcommand{\proj}{\Pi}
\newcommand{\rot}{\mathcal{R}}

\newcommand{\Dc}{\mathcal{D}}

\newcommand{\Hc}{\mathcal{H}}

\newcommand{\Lc}{\mathcal{L}}

\newcommand{\Sc}{\mathcal{S}}

\usepackage{tcolorbox}
\usepackage{mdframed}
\usepackage{lipsum}

\newtheorem{remark}{Remark}[section]

\newcommand{\eqdef}{\ensuremath{\stackrel{\mbox{\upshape\tiny def.}}{=}}}

\graphicspath{{./images/}}

\title{\textsc{Ice-Tide}: Implicit Cryo-ET Imaging and Deformation Estimation}
\author[1,*]{Valentin Debarnot}
\author[1]{Vinith Kishore}
\author[2]{Ricardo D. Righetto}
\author[1,3]{Ivan Dokmani\'c}
\affil[1]{Department of Mathematics and Computer Science, University of Basel, 4051 Basel, Switzerland}
\affil[2]{Biozentrum, University of Basel, 4056 Basel, Switzerland}
\affil[3]{Department of Electrical and Computer Engineering, University of Illinois at Urbana-Champaign, Urbana, IL 61801}
\affil[*]{Correspondence to: \texttt{valentin.debarnot@unibas.ch}}
\usetikzlibrary{patterns}

\usetikzlibrary{arrows.meta}
\usetikzlibrary{decorations.markings}
\newcommand{\revision}[1]{{\leavevmode\color{black}{#1}}}

\begin{document}
	
\maketitle

\begin{abstract}
We introduce ICE-TIDE, a method for cryogenic electron tomography (cryo-ET) that simultaneously aligns observations and reconstructs a high-resolution volume. The alignment of tilt series in cryo-ET is a major problem limiting the resolution of reconstructions. ICE-TIDE relies on an efficient coordinate-based implicit neural representation of the volume which enables it to directly parameterize deformations and align the projections. Furthermore, the implicit network acts as an effective regularizer, allowing for high-quality reconstruction at low signal-to-noise ratios as well as partially restoring the missing wedge information. We compare the performance of ICE-TIDE to existing approaches on realistic simulated volumes where the significant gains in resolution and accuracy of recovering deformations can be precisely evaluated. Finally, we demonstrate ICE-TIDE's ability to perform on experimental data sets.
\footnote{This work was supported by the European Research Council Starting Grant 852821---SWING.
Calculations were performed at sciCORE (\href{http://scicore.unibas.ch/}{http://scicore.unibas.ch/}) scientific computing center at University of Basel. We acknowledge Ben Engel, Lorenz Lamm and Wojciech Wietrzynski for access to data and discussions.}

\end{abstract}

\section{Introduction}

Cryo-electron tomography (cryo-ET) is an imaging technique that enables the study of proteins and other macromolecular structures in their native cellular environment. In cryo-ET, a three-dimensional sample is observed by tilting it around a fixed axis while capturing two-dimensional projections at known viewing directions using the transmission electron microscope (TEM). Recent advances in instrumentation, sample preparation and software have enabled the resolution of protein structures at high resolution \textit{in situ}, leading to a fast recent growth in the popularity of cryo-ET. Notwithstanding, a number of challenges remain for cryo-ET to become a readily accessible and efficient imaging modality.

Due to the sample being a thick slab (100--200 nm), it is not possible to tilt the specimen to very high tilt angles, as this would block the transmission of electrons. In practice, the angular range of a tilt series (TS) typically lies within $-60$ to $+60$ degrees. The incomplete set of viewing directions gives rise to a critical issue known as the missing wedge problem: by the Fourier slice theorem \cite{shkolnisky2012viewing}, there is an unobserved wedge of frequencies in the Fourier transform of the three-dimensional volume. When the tomograms contain distinct patterns such as copies of a macromolecular structure, these subvolumes can be aligned and averaged to compensate for the missing wedge effect, assuming they occur in sufficiently distinct orientations \cite{wan2016cryo}. Alternative approaches which leverage deep learning have recently emerged \cite{liu2022isotropic,wiedemann_deep_2023}.

\revision{The fidelity of three-dimensional cryo-ET reconstructions is further affected by physical perturbations during acquisition. These perturbations manifest in various forms, for example as sample drift between successive acquisitions, leading to misalignment of consecutive projections, or as local deformations in the two-dimensional projections due to beam-induced motion \cite{campbell2012movies,brilot2012beam,zheng2022aretomo}, mainly in the form of a "doming" effect, or charging or the sample which provokes a micro-lensing effect \cite{russo_charge_2018}.
The conventional cryo-ET workflow entails the alignment of tilt series images, followed by the application of standard reconstruction algorithms, such as Iterative Least Squares (ILS), Algebraic Reconstruction Technique (ART) \cite{andersen1984simultaneous}, Simultaneous Iterative Reconstruction Technique (SIRT) \cite{gilbert1972iterative} or, more commonly, Filtered Back-Projection (FBP) \cite{feldkamp1984practical}.}

Not explicitly correcting for the missing wedge still allows a reasonable volume reconstruction despite the typically low signal-to-noise ratios (SNRs), even when using traditional reconstruction algorithms.
On the contrary, not correcting for tilt series deformations  significantly affects the resolution of the final reconstruction, often compounded by the missing wedge. This is demonstrated in a simulated example in Fig. \ref{fig:intro}.

In this paper, we introduce a new cryo-ET reconstruction algorithm, \method{}, and show that it can correct for deformations in the tilt series jointly with accurately reconstructing the 3D volume density.

\def\sz{3.9cm}
\def\ps{0.23}
\begin{figure*}
	\centering
	\begin{subfigure}[t]{\ps\textwidth}
	    \centering
		\includegraphics[height=\sz]{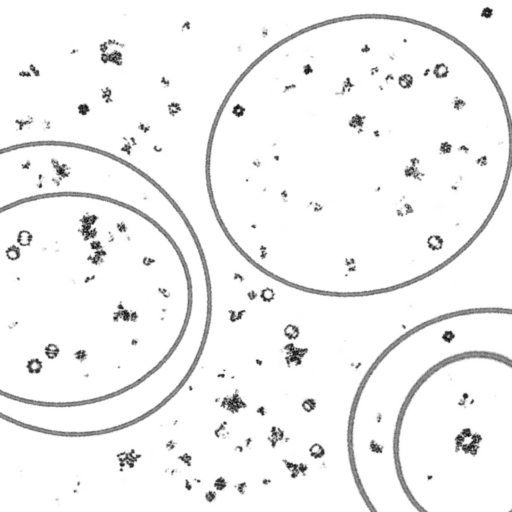}
		\caption*{a) Ground truth}
	\end{subfigure}\hfill
	\begin{subfigure}[t]{\ps\textwidth}
	    \centering
		\includegraphics[height=\sz]{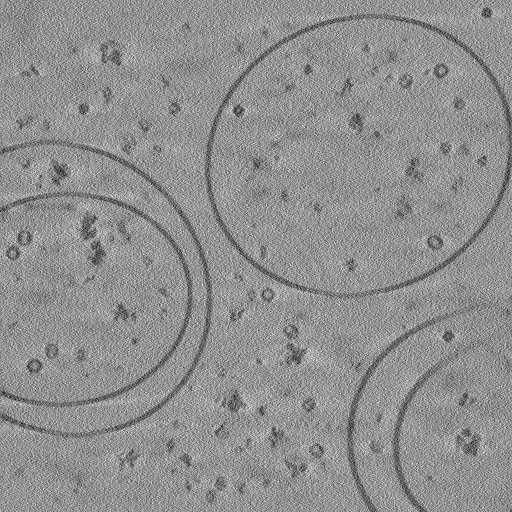}
		\caption*{b) Reconstructed from undeformed TS using FBP}
	\end{subfigure}\hfill
	\begin{subfigure}[t]{\ps\textwidth}
	    \centering
		\includegraphics[height=\sz]{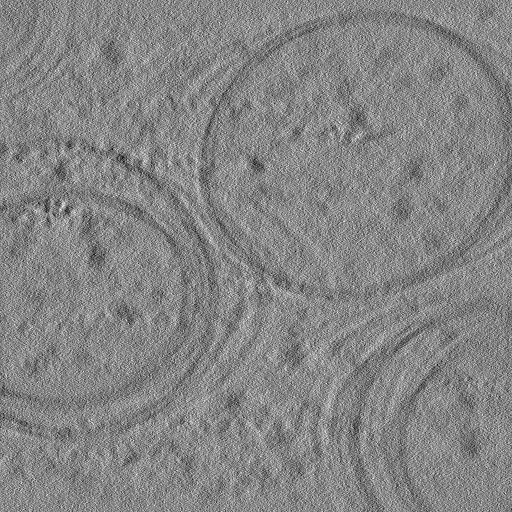}
		\caption*{c) Reconstructed from deformed TS using FBP}
	\end{subfigure}
\caption{Slices through a tomogram reconstruction.}\label{fig:intro}
\end{figure*}

\subsection{Contributions}
We propose a pipeline to reconstruct three-dimensional volumes in cryo-ET from raw tilt series. The originality of our approach lies in that it jointly estimates the volume density and the deformation parameters degrading the tilt series.
The main features of our approach are the following:
\begin{itemize}
	\item 
	We propose an end-to-end \emph{self-supervised} pipeline that directly processes the unaligned tilt-series obtained in tomography to predict the reconstructed tomogram.
	We show that it allows us to sidestep the usual combination of multiple software packages in the reconstruction pipeline and yet leads to comparable or better quality reconstructions.
    \item 
    We leverage an \textit{implicit neural network} to estimate the volume density and its physical model to capture rigid and local deformations.
	\item 
	We show on various realistic simulations the strengths and limitations of our approach. In particular, \emph{\method{} achieves the best possible} reconstruction among methods that are not compensating for the missing wedge. 
	We also explore the limits of our approach in terms of noise and running time.
    \item We illustrate the benefits of ICE-TIDE by reconstructing an experimentally obtained unaligned tilt series. We show that we estimate a tomogram comparable to that obtained using several advanced software packages.
\end{itemize}

All results in this paper are fully reproducible; the open source code is available online.\footnote{\href{https://github.com/swing-research/Implicit-Cryo-Electron-Tomography}{https://github.com/swing-research/Implicit-Cryo-Electron-Tomography}}

\subsection{Related work}
Cryo-ET has been growing as the main imaging modality for reconstructing biological macromolecules with sub-nanometer accuracy in their cellular context. However, the quality of reconstructions in cryo-ET is highly dependent on the expertise of the user operating a complex computational data processing pipeline.
There are numerous software packages  for single-particle cryo-EM, but only a few are specifically tailored for cryo-ET \cite{kremer1996computer, chen_complete_2019, de2022scipiontomo, pyle_current_2021}.
While similarities between the modalities allow certain methods to be applicable in both contexts,  techniques that leverage the unique characteristics of cryo-ET acquisition models tend to yield better reconstructions; we show this empirically in Section \ref{sec:experiments}.

Our approach relies on a careful modeling of the acquisition, in particular regarding the estimation of the latent variables that encode the TS deformation. These parameters are determined through stochastic gradient descent (SGD). We extend and adapt the conceptual framework introduced in a series of earlier works by some of the authors \cite{gupta2022differentiable,debarnot2022joint}.
SGD has recently received considerable attention in single-particle cryo-EM since being leveraged in the cryoSPARC software package \cite{punjani2017cryosparc}. As argued there, it effectively addresses the challenges put forward by the non-convexity of joint pose and density estimation, even starting from random initial configurations.

\subsubsection{Cryo-ET data acquisition, processing and reconstruction}
The first step of the cryo-ET pipeline is to collect projection images of the three-dimensional specimen at different tilt angles. The vitrified sample can only tolerate a finite dose of electrons without being severely damaged. This  results in very low signal-to-noise ratios (SNR).
Multiple short-exposure snapshots are recorded at a high frame rate for each viewing angle via direct electron detectors \cite{li2013electron}. The frames of each movie, despite having extremely low SNR, can be aligned and averaged, compensating for sample drift in a given specimen orientation \cite{Zheng2017}. The aligned frames can then be averaged with exposure weighting to create a projection image \cite{Grant2015}.  The collection of projection images from all the different viewing angles makes up the so-called tilt series.

Furthermore, the contrast transfer function (CTF) defines the frequency-dependent attenuation and phase reversal affecting each observed image. The CTF is a defocus-dependent oscillatory function and arises from the optical elements in the transmission electron microscope as well as the specimen's imperfect placement on the focal plane. Despite the target defocus being defined by the microscope operator, the effective defocus needs to be measured for every image in the tilt series by fitting the CTF function to the image's power spectrum. It can be estimated using existing software such as CTFFIND4 \cite{rohou2015ctffind4} or Warp \cite{tegunov2019real}.


Among the main challenges encountered in the cryo-ET reconstruction pipeline are the substantial sample drift and deformations between successive tilts caused by the electron beam. Registering the images in a tilt series and correcting for the distortions is referred to as \textit{tilt series alignment}. Several tools exist for this step \cite{mastronarde2017automated,chen_complete_2019,fernandez_tomoalign_2021}; the Etomo graphical user interface implemented in IMOD \cite{mastronarde2017automated} is a popular choice. These methods rely on sophisticated local tracking and correlation of features between consecutive images. Gold fiducial markers can be used to track the successive images in the tilt series and derive an alignment model for registration. We focus on the more general case where such markers are not available, which is the case for \textit{in situ} cryo-ET.

A popular recent package which addresses the alignment issue in cryo-ET is AreTomo \cite{zheng2022aretomo}. AreTomo aligns the tilt-series by estimating shift, in-plane rotations and local deformations. The estimated tomogram is then obtained by applying the FBP algorithm on the aligned tilt-series. 
Local deformations are estimated by tracking local features inside patches; that is to say, without fiducials. AreTomo runs on graphics processing units (GPU) and runs automatically with a minimal set of parameters provided by the user.
Similar to our approach, \cite{bogensperger2022joint} propose to jointly estimate the tomogram and the tilt series shift deformations. They however rely on grid-based representation of the volume and estimate only the shifts using a primal--dual algorithm. Here, we are interested in estimating the more complex local deformations.

Finally, once the tilt series is aligned to a satisfactory degree, most cryo-ET software resorts to standard reconstruction algorithms for the ray transform, including the FBP \cite{feldkamp1984practical} or SART \cite{andersen1984simultaneous}. The choice between these algorithms typically hinges on the desired balance between reconstruction quality and computation time.
Due to the very low SNRs in cryo-ET, denoising is another key step in processing the reconstructed volumes.  Techniques  based on the \textit{noise2noise} framework \cite{lehtinen2018noise2noise} enable high-quality denoising and contrast restoration \cite{buchholz2019cryo, liu2022isotropic, wiedemann_deep_2023}. Alternatively, traditional linear methods such as Wiener filtering can be employed \cite{penczek_image_2010, verbeke_self_2024}.

\subsubsection{Implicit representation}
Neural Radiance Field (NeRF) models \cite{mildenhall2021nerf} have gained importance in computer graphics for representing large 3D scenes, but also for solving high-dimensional inverse problems involving partial differential equations \cite{khorashadizadeh2022funknn,vlavsic2022implicit,shi2023harpa,sun2021coil}. They define 3D objects using implicit neural networks, and then consistently synthesize 2D scenes that correspond to the observations.
NeRF models generally exhibit a lower number of parameters compared to grid-based volume representations, making them suitable for representing high-resolution details while remaining relatively easy to train. In this paper, we employ a recent extension \cite{muller2022instant}, which introduces multi-resolution spatial encoding, significantly improving training efficiency. This is particularly beneficial in 3D tomography applications, where data size can be extremely large. Note that NeRF models have shown promise in representing biological volumes, as evidenced in scanning transmission electron microscopy (STEM) \cite{kniesel2022clean}, and can be combined with deformation estimation, as demonstrated in Nerfies \cite{park2021nerfies}. 

\section{Methods}

\subsection{Image formation model}

\begin{figure*}
\centering
\includegraphics[height=9cm]{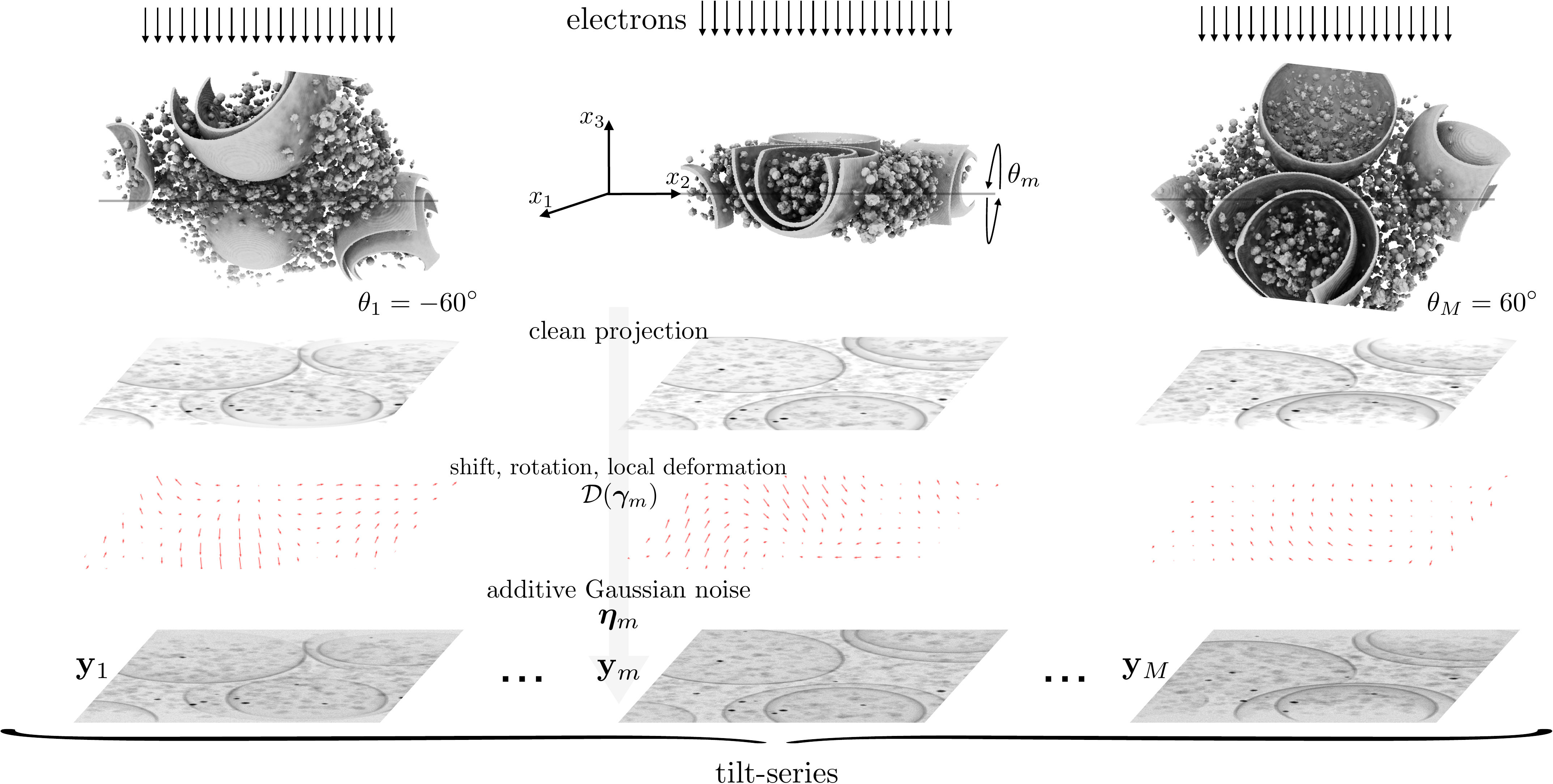}
\caption{Illustration of the cryo-ET acquisition process.} \label{fig:cryoET_process}
\end{figure*}

Let $\rho : \R^3 \to \R_+$ be a bounded, non-negative volumetric density. The cryo-ET forward model can be formally described as the observation of a noisy tilt series composed of $M$ discrete images of size\footnote{For notational simplicity we only consider square projections, but non-square tilt series are easily handled.} $N \times N$, 
\begin{align}
	\yb_{m} = \Hc_m\Dc(\gammab_m^\star)\proj \rot_{\textrm{tilt}}(\theta_m) \rho + \etab_m, \quad m = 1, \ldots, M, \label{eq:cryo-ET_formulation}
\end{align}
where $\rot_{\textrm{tilt}}(\theta_m)$ denotes the rotation (tilt) by an angle $\theta_m$ around the second axis, such that $(\rot_{\textrm{tilt}}(\theta_m)\rho)(\xb) = \rho(\Rb_{\textrm{tilt}}(\alpha)\xb)$ for all $\xb\in\R^3$ with 
$$\Rb_{\textrm{tilt}}(\alpha) = \begin{pmatrix}
	\cos(\alpha) & 0 & \sin(\alpha)\\
    0 & 1& 0 \\ 
	-\sin(\alpha) & 0 & \cos(\alpha)
\end{pmatrix}.$$
The integral projection along the third axis, $\Pi$, is defined via 
$$ (\Pi u)(x_1,x_2) = \int_{-\infty}^\infty u(x_1,x_2,x_3)\mathrm{d}x_3.$$
Finally, $\Dc(\gammab_m^\star)$ denotes the $m$-th deformation affecting the tilt series, $\Hc_m$ models the discretization into a finite set of $N\times N$ pixels, and $\etab_m$ is independent additive random noise. This cryo-ET image formation process is illustrated in Fig. \ref{fig:cryoET_process}.
In cryo-ET, it is reasonable to assume that the CTF and the viewing directions are known. For simplicity, we will ignore the effect of the CTF. It thus remains to estimate the volume density $\rho$ and the deformations $\Dc(\gammab_m^\star)$ based on the observations $\{\yb_m\}_{m=1}^M$.

To summarize, an accurate reconstruction of the unknown density is hindered by three main challenges: the missing wedge due to the fact that the angles $\theta_m$ do not cover the full range, the presence of the \emph{unknown} deformations $\Dc(\gammab_m)$ and the very low SNR.

\subsection{Deformation model}

The mechanical stage drift corresponds to \emph{global deformations}, namely shifts and in-plane rotations~\cite{mastronarde2007fiducial} of the two-dimensional images of the tilt series. 
Unfortunately, the sample is also deformed locally by the electron beam; unlike the mechanical stage drift, this cannot be modeled by rigid transformations alone. 
We propose to decompose the deformation operator $\Dc$ into three elementary operators: a shift $\Sc$, an in-plane rotation $\rot$ and a continuous local deformation $\Phi$.
The final deformation model reads 
\begin{equation}
	\label{eq:deformations}
	\Dc(\gammab) = \Phi(\Cb) \rot(\alpha) \Sc(\taub),
\end{equation}
with parameters $\gammab = \{\Cb, \alpha, \taub\}$ to be estimated.
Letting $u : \R^2 \to \R$ denote a 2-dimensional continuous image, the action of the three components of the deformation model is given as follows:
\paragraph{The shift operator $\mathcal{S}(\taub)$} parameterized by the translation vector $\taub \in\R^2$:
\begin{equation}\label{eq:shift_def}
	\Sc(\taub) u: \xb \mapsto u(\xb-\taub).
\end{equation}

\paragraph{The in-plane rotation operator $\rot(\alpha)$} parameterized by the angles of rotation $\alpha \in [0,2\pi [$:
\begin{equation}\label{eq:rot_def}
	\rot(\alpha) u: \xb \mapsto u\left( \Rb_\alpha \xb\right),
\end{equation}
where $\Rb_\alpha = \begin{pmatrix}
	\cos(\alpha) & -\sin(\alpha)\\
	\sin(\alpha) & \cos(\alpha)
\end{pmatrix}$ is the 2D rotation matrix.

\paragraph{Local continuous deformation} we define a continuous deformation operator $\Phi(\Cb)$ acting on projection images as
\begin{equation}\label{eq:local_def}
	\Phi(\Cb)  y: \xb \mapsto y\left( \xb + \phi(\Cb)(\xb)  \right),
\end{equation}
where $\phi(\Cb):\R^2 \to \R^2$ is a continuous function defining the local displacement at any given position of the field of view.
In this paper, we assume that $\phi(\Cb)$ is defined via an interpolation of a discrete displacement field.
We define $N_1 \times N_2$ displacement vectors at fixed  \emph{control points}. The amplitude and directions of these displacement vectors are the parameters $\Cb$ defining the function $\phi$. There are then $2\times N_1 \times N_2$ parameters.
We use either bilinear or bicubic interpolation to obtain the displacement at arbitrary continuous coordinates; see Fig. \ref{fig:zoom_deformations} for an example.
A possible alternative is to use an implicit neural network to model the deformation field.

\begin{remark}
	We model deformations as an operator acting on the two-dimensional projections of the tilt series. This simplifies the estimation compared to deformation models defined on the volume space while still being sufficiently expressive to approximate volumetric deformations in cryo-ET as shown on real data reconstruction, see Fig. \ref{fig:tkviui}.
\end{remark}
\begin{remark}
	In order to be as general as possible, we do not assume any temporal or sequential relationship between the parameters coefficients $\{\gammab_m\}_{m=1}^M$. Enforcing such a relationship could regularize  the parameters' estimation in challenging settings (e.g., with high noise). Such prior assumptions can be incorporated using a specifically designed regularization term while solving Problem \eqref{eq:opt}.
\end{remark}
\begin{remark}
	We empirically observe much better performance when modeling the three classes of deformations separately; a similar observation has already been made for Nerfies \cite{park2021nerfies}. Unlike Nerfies, we use separate parameterization to capture local deformations from one image to the next, as we do not assume a continuous deformations between them. 
\end{remark}

\subsection{Volume density}
All deformation operators are naturally defined as acting on the coordinate space of the volume density. For this reason, it is crucial to have a representation that allows a fast implementation of these deformations.

We thus represent the volume $\rho$ by an implicit neural representation $ \Vb(\psib): \R^3 \to \R$, parameterized by $\psib$. We use the multi-resolution hash encoding described in \cite{muller2022instant} as positional encoding. It defines multiple grids of features at different resolutions and uses a continuous interpolation of these features as input to a multi-layer perceptron (MLP). The interpolated feature map contains a rich representation of the query position. 

This representation further facilitates the numerical approximation of density integrals along rays in the computation of the forward imaging model \eqref{eq:cryo-ET_formulation}; with fixed grid-based representations one must usually resort to interpolation.

\revision{Empirically, we find that neural field representation of the volume not only offers the advantages previously mentioned but also serves as an excellent prior for addressing the missing wedge problem. The strength of our approach lies in its ability to estimate both volume density and deformations simultaneously, all within a conceptually simple toolbox.
The reconstruction obtained with \method{} can undoubtedly be further improved by using missing wedge correction or denoising tools such as Isonet \cite{liu2022isotropic} or Cryo-CARE \cite{buchholz2019cryo}.
}

\subsection{Estimating the parameters}
We estimate the parameters $$\{\gammab_m = (\Cb_m,\alpha_m,\taub_m)\}_{m=1}^M \quad \text{and} \quad \psib$$  jointly by minimizing the mean absolute error between the observed data and and the tilt series computed from our recontruction, corrected with estimated deformations,
\begin{equation}\label{eq:data}
	\Lc_\mathrm{data}\left( \gammab_m,\psib \right) \eqdef \left\|\yb_m - \Hc_m\boldsymbol{D}(\gammab_m)(\Pb(\Rb_{\theta_m}(\boldsymbol{\Vb(\psib)})) \right\|_1.
\end{equation}
The use of the $\ell^1$-norm instead of the statistically optimal (for i.i.d. Gaussian noise) $\ell^2$-norm is common in deep learning for imaging; empirically, it leads to faster convergence and better quality of reconstruction.

\begin{remark}
The formulation \eqref{eq:data} allows mini-batch optimization in the pixel space, contrary to the formulation presented in \cite{debarnot2022joint}. This feature is key to deal with the large size of tilt series collected in cryo-ET experiments.
\end{remark}

In this paper, we assume that the deformation operators are ``small'' (close to identity) which is indeed the case in cryo-ET.
This prior is enforced by penalizing the size of the parameters of the deformation operators by minimizing
\begin{equation}
	\label{eq:reg_amplitude}
    \begin{split}
	\Lc_\mathrm{deform}(\gammab_m) &\eqdef  \lambda_1\min(|\alpha_m|, |2\pi-\alpha_m|)\\
        &+ \lambda_2\left\|\taub_m\right\|_2 + \lambda_3\left\| \Phib(\Cb_m)\right\|,
    \end{split}
\end{equation}
where $\lambda_1, \lambda_2, \lambda_3\geq 0$ are tunable parameters and
$$\left\| \Phib(\Cb_m)\right\| \eqdef \sqrt{\sum_{i=1}^P  \phi(C_m)(\xb_i)^2 },$$
where $\{\xb_i\}_{i=1}^P$ is a uniform mesh of the field of view with $P$ points.

The overall optimization problem to be solved reads 
\begin{equation}\label{eq:opt}
	\min_{\{\gammab_m\}_{m=1}^M, \psib}  \sum_{m=1}^M w_m\Lc_\mathrm{data}\left(\gammab_m,\psib \right) + \Lc_\mathrm{deform}(\gammab_m),
\end{equation}
where $w_m\in\R$ are chosen manually and can take into account different electron dose scheme depending on the tilt angle.
\revision{The optimization problem \eqref{eq:opt} being non-convex, we rely on the stochastic mini-batch gradient descent using the \textit{Adam} algorithm implemented in \textit{Pytorch} \cite{paszke2017automatic}. All the gradients are computed using automatic differentiation and the learning rate is chosen in order to ensure decrease of the cost function on a randomly chosen subset of the observations. We stop the iterative optimization process when the loss function stop decreasing or when the estimated volume is no longer refined.
This approach allows the partitioning of the observations and the cost function on small batch that can be store on GPUs ith reasonably low memory, which is crucial to process the large amount of data typically acquired in cryo-ET.}

We do not impose any explicit regularization, such as the total variation, on the volume density. We observe that this is not needed as the favorable inductive bias of the implicit neural representation yields reconstructions with suppressed noise and enhanced detail. Similar empirical observations have been made elsewhere \cite{kim2022zero} and have been theoretically connected with the spectral properties of neural network architectures \cite{chakrabarty2019spectral, tachella2021neural}. 
\revision{They find that} implicit regularization even partially restores the missing wedge information, as shown in Fig. \ref{fig:fourier}.

\section{Results}\label{sec:experiments}

\subsection{Simulated dataset}

We assess our approach on a realistic simulated dataset which allows us to precisely quantify the quality of the reconstruction and the estimated deformations.

We use the SHREC 2021 dataset \cite{shrec2021}, which consists of tomograms sampled at  1 nm/voxel and with total size of $512\times 512\times 512$ voxels. 
Each tomogram is composed of different simulated proteins of various sizes and shapes in random orientations.
We use this dataset to obtain an initial realistic tomogram for our experiments. Subsequently, we apply the forward model described by Equation \eqref{eq:cryo-ET_formulation} to generate a tilt series with $61$ projection images with viewing angles varying between $-60$ and $+60$ degrees at 2 degree steps. 
The deformation parameters are sampled independently at random with small amplitude, such that the maximum shift is 12 pixels, the maximum in-plane rotation is 0.01 degrees and the local deformations induce a displacement of at most 2 pixels. The true local deformations are obtained using Model \eqref{eq:local_def}; to avoid an inverse crime, the deformation is defined on $5\times 5$ grid of control points while our algorithm uses a finer grid of $10\times 10$ control points, which, crucially, do not include the true control points as a subset. 
We perturb the tilt series with global and local deformations and add Gaussian noise to obtain a final SNR of 10 dB, where the SNR between a clean signal $\yb$ and the perturbed one $\yb'$ is defined as 
\begin{align}
	\text{SNR}(\yb,\, \yb') = -20 \log_{10} \left(\frac{\|\yb - \yb'\|_2^2}{\|\yb\|_2^2} \right).
\end{align}

Several images from the tilt series are displayed in Fig. \ref{fig:projection_deformaton_example}.
The deformations present can be better appreciated from a video of the tilt series; please see \href{https://github.com/swing-research/Implicit-Cryo-Electron-Tomography/tree/main/videos}{https://github.com/swing-research/Implicit-Cryo-Electron-Tomography/tree/main/videos}.

We measure the quality of the reconstruction using Fourier shell correlation (FSC) \cite{Harauz1986}. The FSC between two discrete volumes $\rhob, \rhob'$ at a radius $r$ of the shell is defined as 

\begin{equation}\label{eq:FSC}
    \textrm{FSC}(\rhob,\rhob')[r] =
    \frac{\sum_{|\xib|= r } \widehat{\rhob}(\xib) \widehat{\rhob}'(\xib)}{\sqrt{\sum_{|\xib|= r } \widehat{\rhob}(\xib)^2\sum_{|\xib'|= r } \widehat{\rhob}'(\xib')^2}},
\end{equation}
where $\widehat{\rhob}$ (resp. $\widehat{\rhob}'$) denotes the three-dimensional Fourier transform of the discrete volume $\rhob$ (resp. $\widehat{\rhob}'$).
The FSC measures the correlation between the estimated and the original volume in a frequency-dependent manner, giving  information on the quantity of details that can be recovered.
\revision{Importantly, after the tilt-series alignment procedure, the reconstructed volumes may be centered on a different reference frame than the ground truth volume. To prevent this shift, which could result in a poor FSC despite the volume retaining all high-resolution features, we first aligned the two volumes using a standard correlation-based registration algorithm.}
Additionally, we report the normalized Correlation Coefficient (CC) between the estimate and the true volume. 
The correlation coefficient has the advantage of summarizing reconstruction quality in a single number, making it easier to compare different experimental settings. 
The correlation coefficient between two discrete volumes $\rhob, \rhob'$ is defined as 
\begin{equation}\label{eq:CC}
\begin{split}
    &\textrm{CC}(\rhob,\rhob') = \frac{\left\langle \rhob, \rhob' \right\rangle}{\|\rhob\|_2\|\rhob'\|_2},
\end{split}
\end{equation}
where $\left\langle \rhob, \rhob' \right\rangle$ defines the scalar product and $\|\rhob\|_2$ defines the $\ell_2$-norm.
Note that FSC and CC are two different metrics and should be understood as complementary.
The estimated volume may sometimes suffer from a global shift which has no influence on the quality of the reconstruction, but which may deteriorate the results of the FSC or CC metrics if not accounted for. For this reason, we systematically align the estimated volumes with the ground truth before reporting the FSC or CC.

\def\sz{4.4cm}
\def\xx{0}
\def\yy{0}
\def\zx{-0.9}
\def\zy{0.3}
\def\ps{0.24}
\begin{figure*}
	\centering
	\begin{subfigure}[t]{\ps\textwidth}
	    \centering
	    \begin{tikzpicture}[spy using outlines={circle,orange,magnification=4,size=1.5cm, connect spies}]
    		\node[ rotate=90] at (0*\xx,-0*\yy) {}; outlines={circle,orange,magnification=2,size=3cm, connect spies}]
    		\node[rotate=0, line width=0.05mm, draw=white] at (0*\xx,-0.*\yy) { \includegraphics[height=\sz]{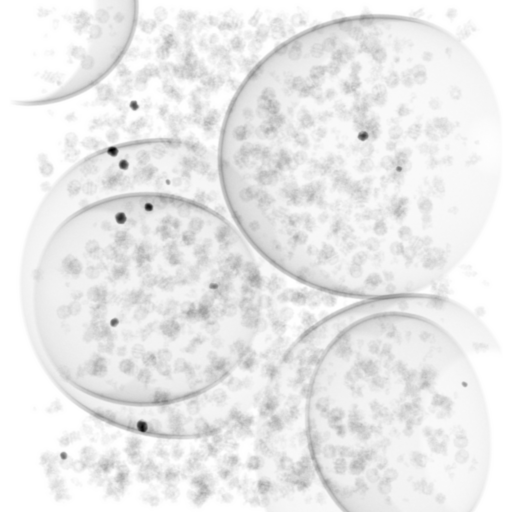}};
    		\spy on (\zx,\zy) in node [left] at (-0.6,-1);
		\end{tikzpicture} 
		\caption*{a) Ground truth}
	\end{subfigure}\hfill
	\begin{subfigure}[t]{\ps\textwidth}
	    \centering
	    \begin{tikzpicture}[spy using outlines={circle,orange,magnification=4,size=1.5cm, connect spies}]
    		\node[ rotate=90] at (0*\xx,-0*\yy) {}; outlines={circle,orange,magnification=2,size=3cm, connect spies}]
    		\node[rotate=0, line width=0.05mm, draw=white] at (0*\xx,-0.*\yy) { \includegraphics[height=\sz]{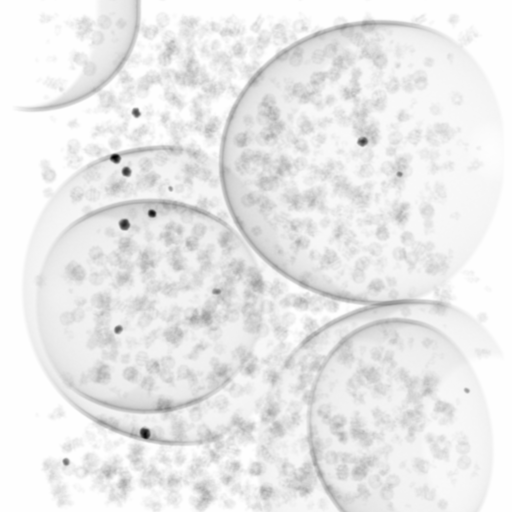}};
    		\spy on (\zx,\zy) in node [left] at (-0.6,-1);
            \draw [orange, decoration={markings,mark=at position 1 with {\arrow[scale=0.4]{>}}}, postaction={decorate},shorten >=0.4pt](-1.03,0.29) -- (-1.0,0.24);
            \draw [orange, decoration={markings,mark=at position 1 with {\arrow[scale=0.4]{>}}}, postaction={decorate},shorten >=0.4pt](-0.9,0.47) -- (-0.9,0.4);
		\end{tikzpicture} 
		\caption*{b) Deformed}
	\end{subfigure}\hfill
	\begin{subfigure}[t]{\ps\textwidth}
	    \centering
	    \begin{tikzpicture}[spy using outlines={circle,orange,magnification=4,size=1.5cm, connect spies}]
    		\node[ rotate=90] at (0*\xx,-0*\yy) {}; outlines={circle,orange,magnification=2,size=3cm, connect spies}]
    		\node[rotate=0, line width=0.05mm, draw=white] at (0*\xx,-0.*\yy) { \includegraphics[height=\sz]{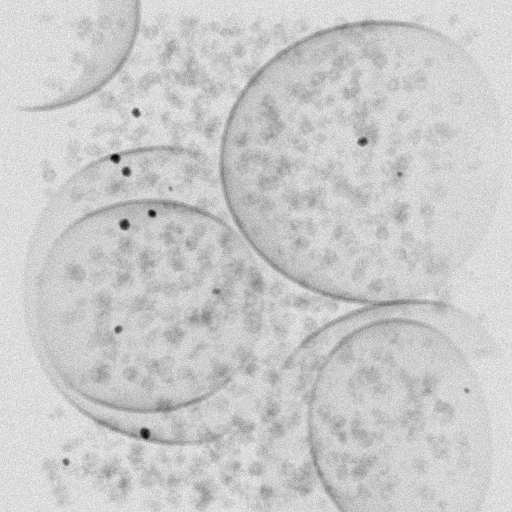}};
    		\spy on (\zx,\zy) in node [left] at (-0.6,-1);
		\end{tikzpicture} 
		\caption*{c) Deformed + noise}
	\end{subfigure}
\caption{Example of simulated projections.}\label{fig:projection_deformaton_example}
\end{figure*}
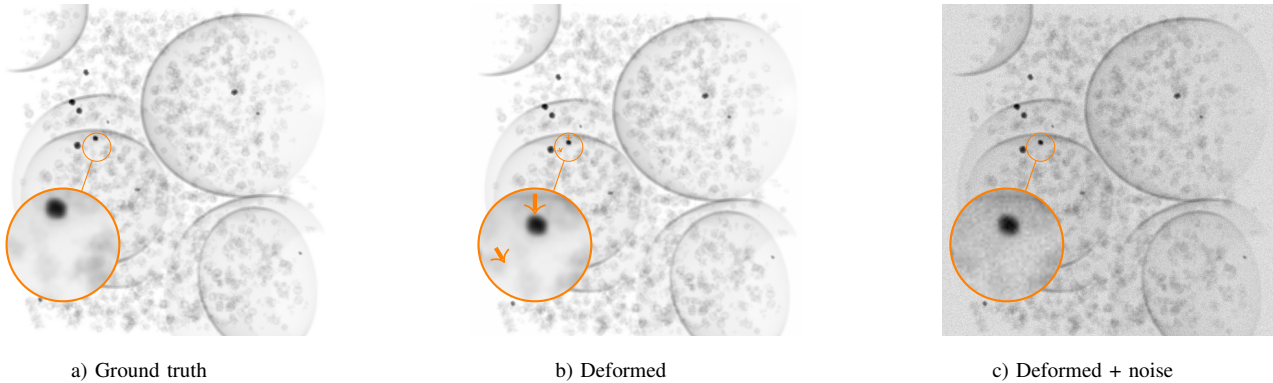

\subsection{Comparison with other common methods}
We compare \method{} with AreTomo \cite{zheng2022aretomo} and Etomo from IMOD \cite{kremer1996computer,mastronarde2017automated}, popular tools for tilt series alignment. Similar to our method, AreTomo estimates shifts, rotations and local deformations. Etomo can align the tilt series interactively using a graphical interface. With Etomo, we interact with the graphical interface in order to select parameters that provide a satisfying result on the volume \textit{model 0} of the SHREC dataset. \revision{Notice that we only use the global alignment feature of Etomo.} In order to perform extensive numerical investigation, we then use the equivalent of Etomo from the command line (\texttt{batchruntomo}), using always the same hyperparameters selected with volume \textit{model 0}.
Hyperparameters used in \method{} can be found in the official repository containing all the following experiments: \href{https://github.com/swing-research/Implicit-Cryo-Electron-Tomography}{https://github.com/swing-research/Implicit-Cryo-Electron-Tomography}.

\revision{
We report the reconstructed volume using FBP and Total Variation (TV) on the projections without deformations but in the presence noise (\textit{FBP w/o def} and \textit{TV w/o def}). \textit{FBP w/o def} represents the best estimation we could hope to achieve while not compensating for the missing wedge. TV regularization is a popular a priori to solve ill-posed inverse problems as it is the case with cryo-ET. We use the Python implementation of the variational approach \cite{chambolle2011first} in Tomopy \cite{gursoy2014tomopy}. \textit{TV w/o def} is a reasonable example of what is possible to achieve while compensating for the missing wedge and \emph{knowing perfectly} the true deformations.
We also report the reconstructed volume using TV applied to the raw (deformed and noisy) projections (\textit{TV w/ def}) in order to show the importance of estimating the deformations even when regularizing the reconstruction problem.  
}

\subsubsection{Relying only on the acquired projections}

The volume produced by solving Problem \eqref{eq:opt} is given by sampling the implicit neural network $\Vb(\psib)$. In practice it is sometimes desired to use a reconstruction obtained only using the tilt series observations and not the output of a deep neural network. Fortunately, our method can be used to retain only the estimated deformations, which have been shown to be highly accurate, to create an aligned version of the tilt series. Consequently, standard reconstruction algorithms like FBP can be used to reconstruct the volume. We report this result as \textit{FBP + ICETIDE} in the following and show that it results in almost the same quality as the reconstruction obtained using FBP on the true undeformed noisy tilt series (which is unavailable in reality). This is especially useful when applying the tilt series alignment on "odd/even" versions of the tilt series required by some denoising methods \cite{lehtinen2018noise2noise,buchholz2019cryo,wiedemann_deep_2023}.

\subsection{Reconstructing a single volume}
We investigate the performance of each method on the first volume of the SHREC dataset, \textit{model 0}. We degrade the tilt-series such that the SNR with the noiseless and deformed projections reaches 10 dB.
We compare the quality of the estimated volumes with the FSC curves in Fig. \ref{fig:fsc}.
We observe that AreTomo and Etomo underpeform. Although AreTomo is slightly above \textit{Etomo}, it is still a long way from what can be achieved with correctly aligned tilt-series (\textit{FBP w/o def}).
\revision{In contrast, ICE-TIDE exhibits strong performance, better than the best FBP reconstruction (\textit{FBP w/o def}). 
Because the volume estimated by ICE-TIDE is not constrained in the missing wedge, it can assume arbitrary values there; again, the inferred values are well-structured and represent the natural volume density. \revision{This difference between the FBP reconstruction and \method{} explains the missing wedge artefacts that are present in Fig. \ref{fig:central_slice_default_experiment}.}
We show the Fourier transform of the central slice of the estimated volume in Fig. \ref{fig:fourier}. We observe that for \revision{the TV} reconstruction, the missing wedge is only poorly filled. 
This confirms our earlier assertion that the inductive bias of the implicit neural network acts as an effective regularizer, well-matched to the spatial statistics of cryo-ET volumes: the estimated volume produced by ICE-TIDE recovers some of the missing wedge, even though it has not been explicitly trained for that. }

In the presence of noise, it is important to stop the iterative resolution of \eqref{eq:opt} early enough in order to not overfit to the noise. This phenomenon is known as early stopping and is key in self-supervised fitting using neural network as in Deep Image Prior \cite{ulyanov2018deep} or \textit{noise2noise} \cite{lehtinen2018noise2noise}. In our experiments, we observe that the overfitting regime appears long after the training loss stabilizes, which provides a convenient early stopping signal.
This interpretation is supported by the visual inspection of orthogonal XYZ slices of reconstructed volumes in Fig. \ref{fig:central_slice_default_experiment}. We observe that the streaking artifacts due to the missing wedge and the discrete number of projections are significantly less pronounced in ICE-TIDE compared to the \revision{FBP and TV reconstructions. }
We also observe certain artifacts in AreTomo reconstructions, where some elements of the volume are stretched along the $z$-axis. This could explain the poorer performance observed on the FSC curves.
We note that the volume reconstructed using TV reconstruction, with only the  deformation parameters being estimated by ICE-TIDE, \revision{closely follows the FSC curve \textit{TV w/o def}}. This can be observed in Fig. \ref{fig:central_slice_default_experiment}, where the densities of the very small structures are similar to those obtained from projections without deformations.

We visually compare the displacement vector fields of the estimated and true local deformations in Fig. \ref{fig:zoom_deformations}. We observe that AreTomo incorrectly estimates the true deformations while ICE-TIDE is able to capture most local deformations accurately.

\def\a{0.7}
\def\lw{0.9pt}
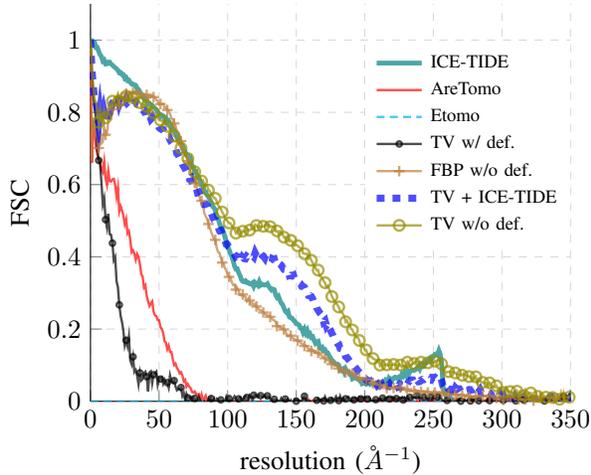
\begin{figure}
	\centering
	\begin{tikzpicture}     
		\begin{axis}[
			width=0.99\linewidth, 
			grid=major, 
			grid style={dashed,gray!30}, 
			xlabel= resolution ($\r{A}^{-1}$),
			ylabel=FSC,
			xtick={0,50,100,150,200,250,300,350},
			xmin = 0, xmax = 350,
			ymin = 0,
			axis x line*=bottom,
			axis y line*=left,
			legend style={at={(1.,0.9)}, legend cell align=left, align=left, draw=none,font=\scriptsize},
			xticklabel style={
				/pgf/number format/fixed,
				/pgf/number format/precision=0,
				/pgf/number format/fixed zerofill
			},
			scaled x ticks=false,
			]
			\addplot[color=teal,line width=2*\lw, opacity=\a] table [x=x, y=icetide, col sep=comma] {images/default_experiment/FSC.csv};
			\addlegendentry{ICE-TIDE}
			\addplot[color=red,line width=\lw, opacity=\a] table [x=x, y=AreTomo_patch1, col sep=comma] {images/default_experiment/FSC.csv};
			\addlegendentry{AreTomo}
			\addplot[color=cyan,line width=\lw, opacity=\a, style=densely dashed] table [x=x, y=ETOMO, col sep=comma] {images/default_experiment/FSC.csv};
			\addlegendentry{Etomo}
			\addplot[color=black,line width=\lw, opacity=\a,mark=o, mark size=1,mark repeat=5,mark phase=7] table [x=x, y=tv, col sep=comma] {images/default_experiment/FSC.csv};
			\addlegendentry{TV w/ def.}
			\addplot[color=brown,line width=\lw, opacity=\a,mark=+, mark size=2,mark repeat=5,mark phase=7] table [x=x, y=FBP_no_deformed, col sep=comma] {images/default_experiment/FSC.csv};
			\addlegendentry{FBP w/o def.}
			\addplot[color=blue,line width=3*\lw, opacity=\a, style=dashed] table [x=x, y=icetide_tv, col sep=comma] {images/default_experiment/FSC.csv};
			\addlegendentry{TV + ICE-TIDE}
			\addplot[color=olive,line width=\lw, opacity=\a,mark=o, mark size=2,mark repeat=5,mark phase=7] table [x=x, y=tv_no_deformed, col sep=comma] {images/default_experiment/FSC.csv};
			\addlegendentry{TV w/o def.}
		\end{axis}   
	\end{tikzpicture}	
	\caption{\revision{Fourier shell correlation with the true volume density. }\label{fig:fsc}} 
\end{figure}

\def\xx{3}
\def\yy{1.5}
\def\zx{-1.3}
\def\zy{1.6}
\def\sz{5.3cm}
\begin{figure*}
    \centering
    \begin{subfigure}[b]{0.5\textwidth}
	\begin{tikzpicture}[spy using outlines={circle,orange,magnification=2,size=1.5cm, connect spies}]    		\node[rotate=0, line width=0.05mm, draw=white] at (0*\xx,-0.*\yy) { \includegraphics[height=\sz]{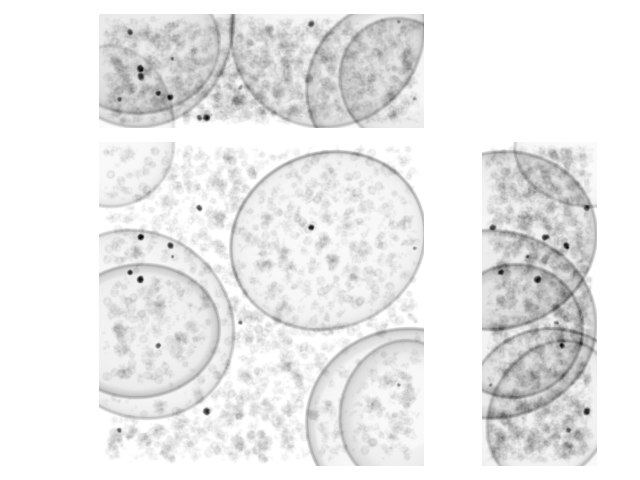}};
		\spy on (\zx,\zy) in node [left] at (\xx,\yy);
	\end{tikzpicture} 
   \caption{Reference volume}
   \label{fig:vol_REF} 
    \end{subfigure}
    \begin{subfigure}[b]{0.49\textwidth}
	\begin{tikzpicture}[spy using outlines={circle,orange,magnification=2,size=1.5cm, connect spies}]    		\node[rotate=0, line width=0.05mm, draw=white] at (0*\xx,-0.*\yy) { \includegraphics[height=\sz]{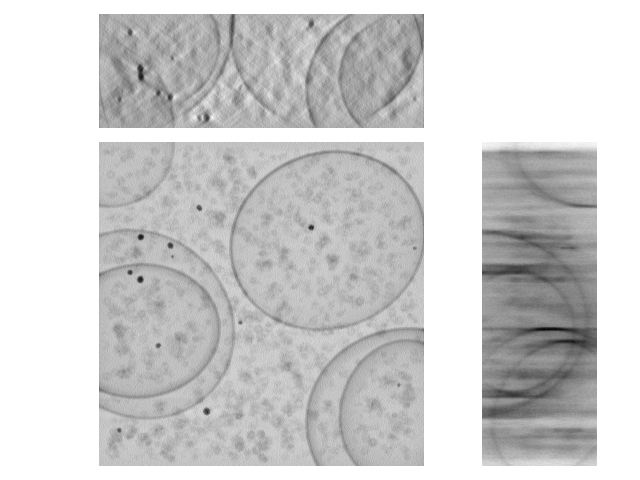}};
		\spy on (\zx,\zy) in node [left] at (\xx,\yy);
	\end{tikzpicture} 
   \caption{TV w/o def. CC: 0.77. }
   \label{fig:FBP_no_def} 
    \end{subfigure}
    \begin{subfigure}[b]{0.49\textwidth}
	\begin{tikzpicture}[spy using outlines={circle,orange,magnification=2,size=1.5cm, connect spies}]    		\node[rotate=0, line width=0.05mm, draw=white] at (0*\xx,-0.*\yy) { \includegraphics[height=\sz]{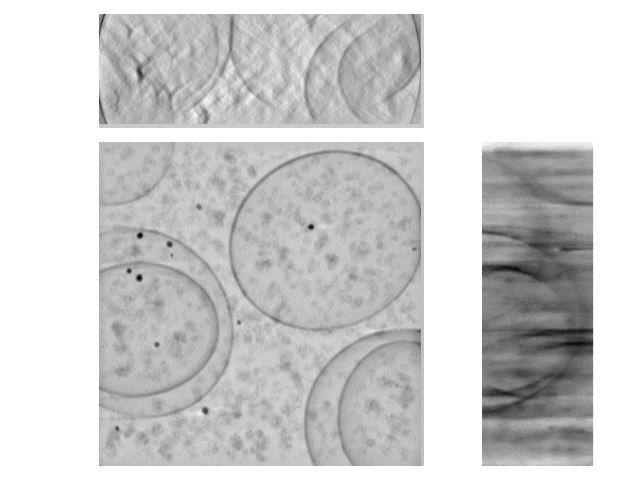}};
		\spy on (\zx,\zy) in node [left] at (\xx,\yy);
	\end{tikzpicture} 
   \caption{AreTomo reconstruction. CC: 0.31.}
   \label{fig:vol_AreTome} 
    \end{subfigure}
    \begin{subfigure}[b]{0.49\textwidth}
	\begin{tikzpicture}[spy using outlines={circle,orange,magnification=2,size=1.5cm, connect spies}]    		\node[rotate=0, line width=0.05mm, draw=white] at (0*\xx,-0.*\yy) { \includegraphics[height=\sz]{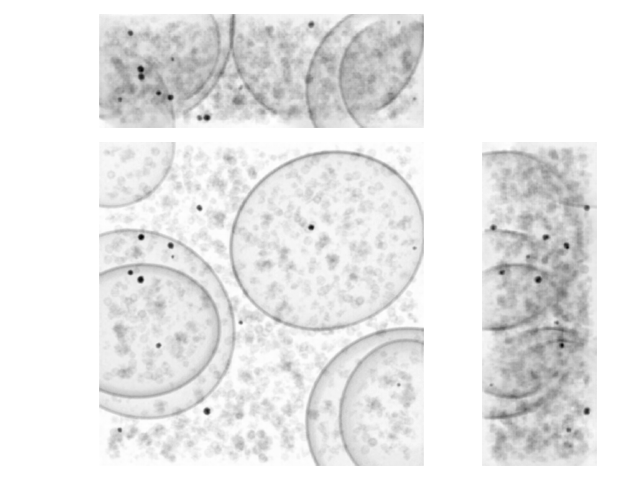}};
		\spy on (\zx,\zy) in node [left] at (\xx,\yy);
	\end{tikzpicture} 
   \caption{ICE-TIDE reconstruction. CC: 0.87.}
   \label{fig:vol_ours} 
    \end{subfigure}
    \caption{Orthogonal slices of different reconstruction of the SHREC \textit{model 0} dataset at SNR 10dB. Contrast has been stretched for better visualization. \revision{Gold fiducials are reconstructed without artefacts by ICE-TIDE.}}
    \label{fig:central_slice_default_experiment}
\end{figure*}

\def\wid{6.8cm}
\def\hs{-1cm}
\def\xx{6.6}
\def\yy{-4.8}
\def\zx{-0.5}
\def\zy{-0.5}
\begin{figure*}
	\hspace{0cm}
	\centering
	\begin{tikzpicture}[spy using outlines={circle,orange,magnification=2,size=1.5cm, connect spies}]
		\node[text width=2cm,text centered,minimum width=2cm,minimum height=2cm, rotate=90] at (-3.6,0) {ICE-TIDE};
		\node[rotate=0] at (0*\xx,0.*\yy) { \includegraphics[width=\wid]{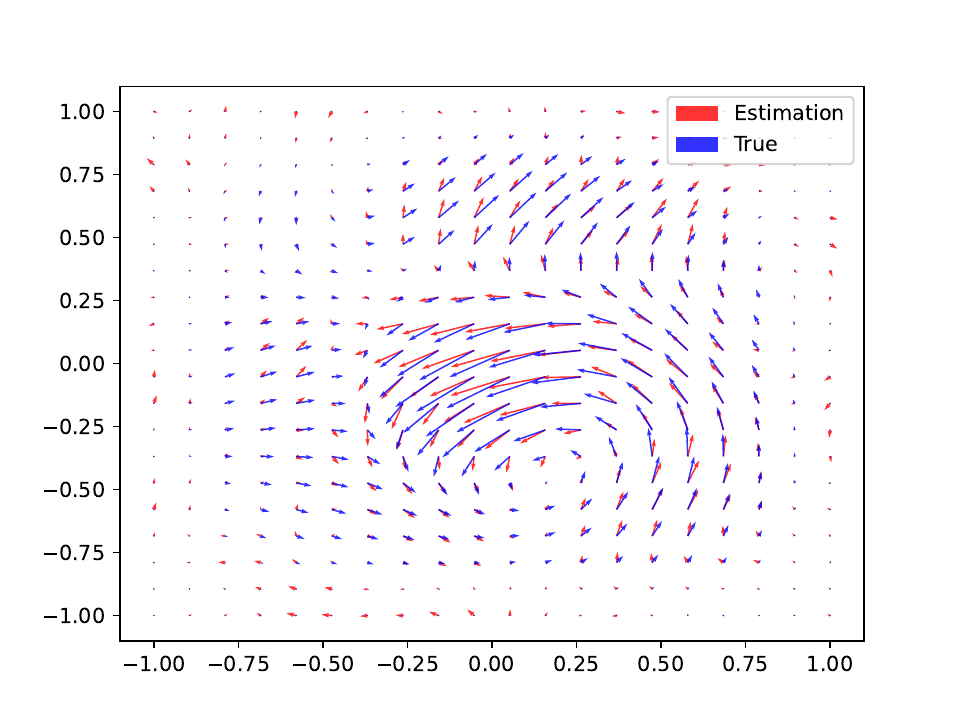}};
		\node[rotate=0] at (1*\xx,0.*\yy) { \includegraphics[width=\wid]{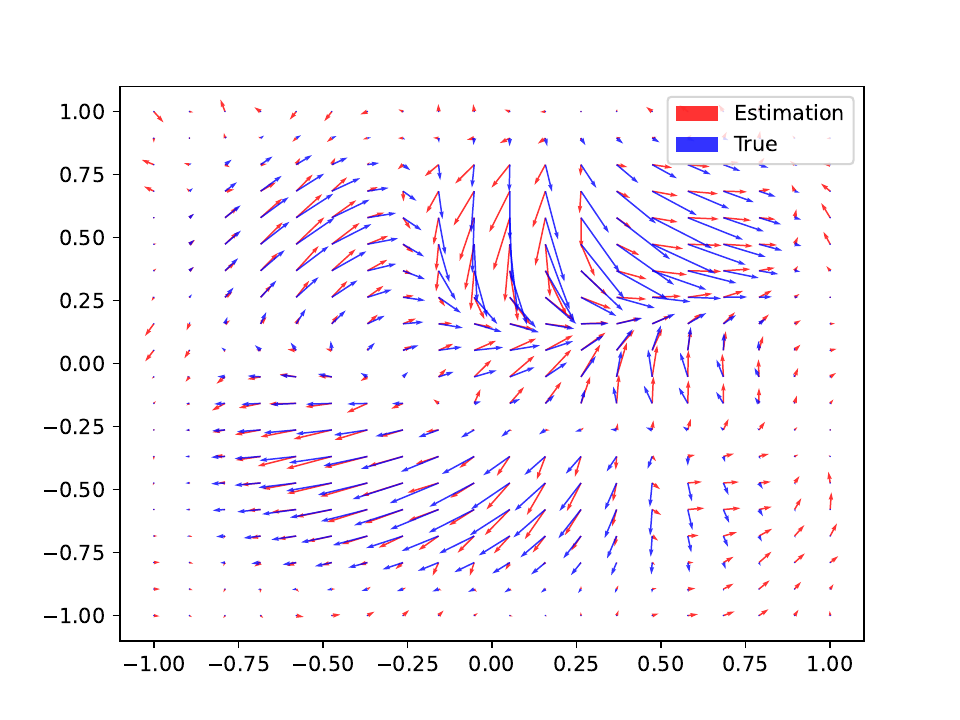}};
		\node[text width=2cm,text centered,minimum width=2cm,minimum height=2cm, rotate=90] at (-3.6,\yy) {AreTomo};
		\node[rotate=0] at (0*\xx,1*\yy) { \includegraphics[width=\wid]{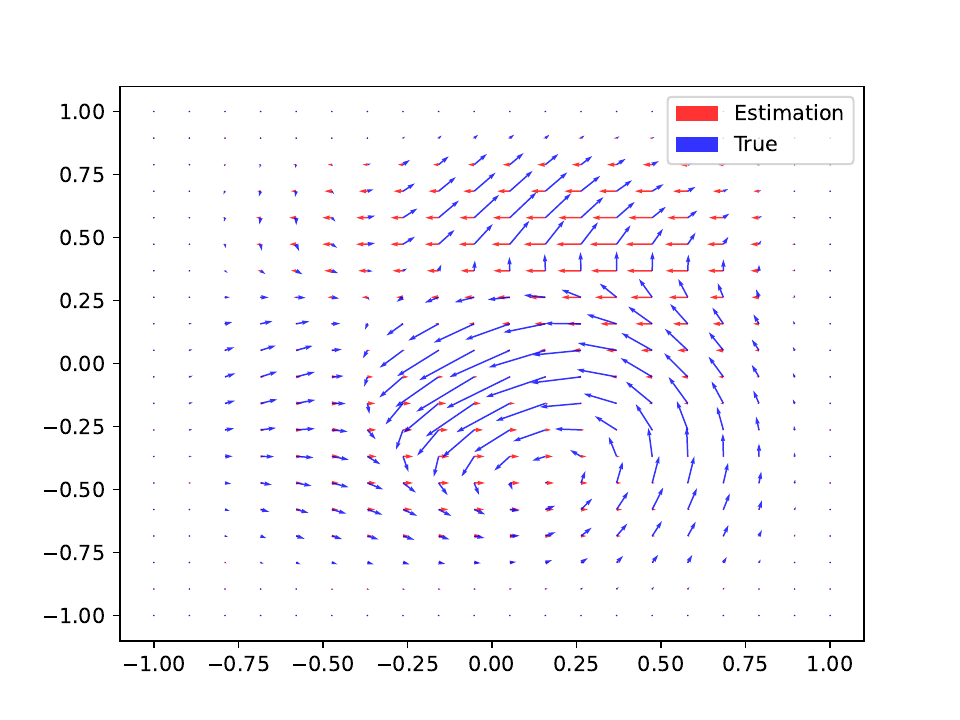}};
		\node[rotate=0] at (1*\xx,1*\yy) { \includegraphics[width=\wid]{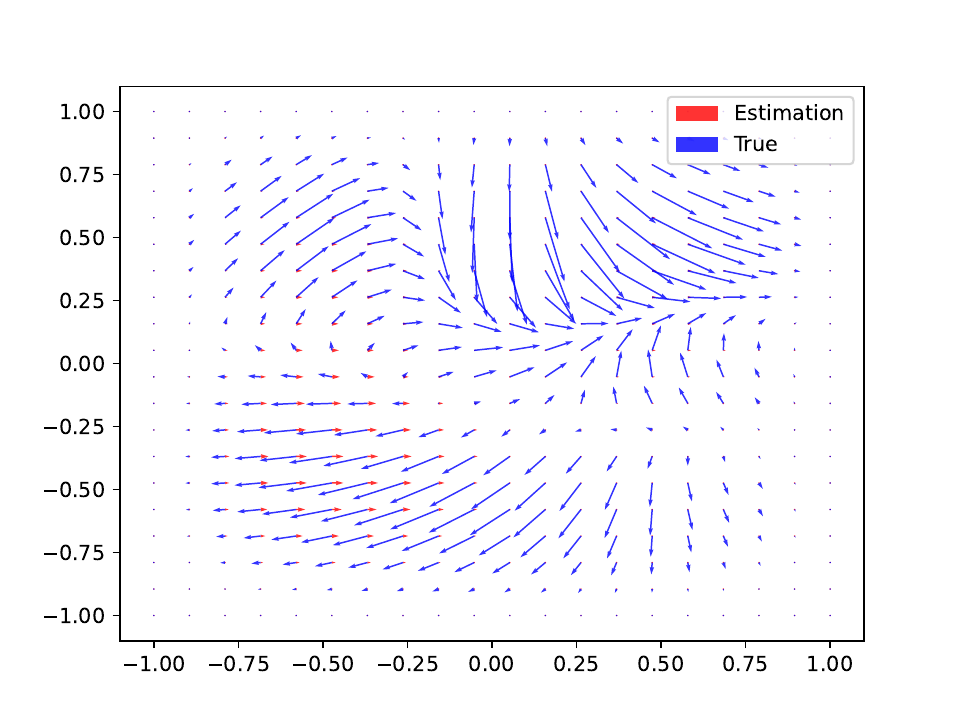}};
		
		\spy on (\zx,\zy) in node [left] at (-0.8,1);
		\spy on (\zx+\xx,\zy) in node [left] at (-0.8+\xx,1);
		
		\spy on (\zx,\zy+\yy) in node [left] at (-0.8,1+\yy);
		\spy on (\zx+\xx,\zy+\yy) in node [left] at (-0.8+\xx,1+\yy);
	\end{tikzpicture} 
	\caption{Two consecutive local deformation maps on the full field of view. Top row: ICE-TIDE; Bottom row: AreTomo. For better visualization, the size of the arrows represents 10 times the actual displacement.\label{fig:zoom_deformations}} 
\end{figure*}

\def\ww{5cm}
\def\ps{0.5}
\def\sz{2.5cm}
\def\xx{0}
\def\yy{0}
\def\zx{-0.}
\def\zy{0.}
\begin{figure}
	\begin{subfigure}[t]{\ps\textwidth}
	    \centering
	    \begin{tikzpicture}[spy using outlines={circle,orange,magnification=3,size=2cm, connect spies}]
    		\node[rotate=90] at (0*\xx,-0*\yy) {}; outlines={circle,orange,magnification=2,size=3cm, connect spies}]
    		\node[rotate=0, line width=0.05mm, draw=white] at (0*\xx,-0.*\yy) { \includegraphics[height=\sz]{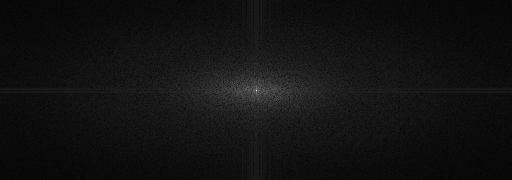}};
    		\spy on (\zx,\zy) in node [left] at (-1.6,0);
		\end{tikzpicture} 
		\caption*{a) Ground truth}
	\end{subfigure}\par\medskip
	\begin{subfigure}[t]{\ps\textwidth}
	    \centering
	    \begin{tikzpicture}[spy using outlines={circle,orange,magnification=3,size=2cm, connect spies}]
    		\node[ rotate=90] at (0*\xx,-0*\yy) {}; outlines={circle,orange,magnification=2,size=3cm, connect spies}]
    		\node[rotate=0, line width=0.05mm, draw=white] at (0*\xx,-0.*\yy) { \includegraphics[height=\sz]{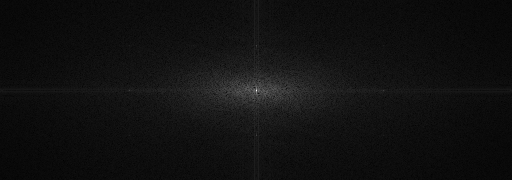}};
    		\spy on (\zx,\zy) in node [left] at (-1.6,0);
		\end{tikzpicture} 
		\caption*{b) ICE-TIDE}
	\end{subfigure}\par\medskip
	\begin{subfigure}[t]{\ps\textwidth}
	    \centering
	    \begin{tikzpicture}[spy using outlines={circle,orange,magnification=3,size=2cm, connect spies}]
    		\node[ rotate=90] at (0*\xx,-0*\yy) {}; outlines={circle,orange,magnification=2,size=3cm, connect spies}]
    		\node[rotate=0, line width=0.05mm, draw=white] at (0*\xx,-0.*\yy) { \includegraphics[height=\sz]{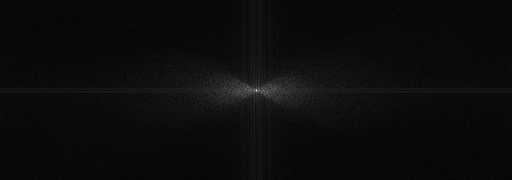}};
    		\spy on (\zx,\zy) in node [left] at (-1.6,0);
		\end{tikzpicture} 
		\caption*{c) \revision{TV w/o def}}
	\end{subfigure}
\caption{\revision{Magnitude of the Fourier transform of the central slice of the SHREC \textit{model 0} volume reconstructed at SNR 10dB. Contrast has been stretched for better visualization.} }\label{fig:fourier}
\end{figure}

\subsection{Time and memory}

One key strength of ICE-TIDE is that it can perform batch sampling and then trade-off memory for time without downsampling the tilt series data. As a result, it can be run on low-resource hardware, with correspondingly longer compute time for a fixed reconstruction accuracy. In Table \ref{tab:time-memory}, we compare the time and memory usage with AreTomo and Etomo.
AreTomo has an option to perform only global alignment and disregard the local patch based alignment scheme, which consumes a significant amount of running time. In \method{}, there is no significant difference in terms of computation time between estimating or not estimating local deformations.
ICE-TIDE simultaneously estimates the original volume and the deformation, which requires a considerable amount of processing time. The additional reconstruction time is compensated by the quality of deformation and volume estimation, but also by the fact that large tilt series can be processed without downsampling. A reconstruction can be obtained in less than 20 minutes on a computer with CPU Intel i9 and GPU NVIDIA GeForce RTX 3090.

\begin{table}
	\begin{center}
		\begin{tabular}{ m{1.5cm}|m{0.8cm}|m{1.4cm}|m{1.2cm}|m{0.8cm}} 
			\toprule
			& ICE-TIDE & AreTomo only global & AreTomo $4\times 4$ patches & Etomo (batch)\\ 
			\hline
			Time (s) & \centering{1137} & \centering{54}  & \centering{2364} &  {\centering{\textbf{25}}} \\ 
   			Memory (MB)& \centering{1058} & \centering{450} & \centering{461} & {\centering{\textbf{381}}} \\
			\bottomrule
		\end{tabular}
		\caption{Time and memory consumption for the different methods to reconstruct the SHREC \textit{model 0} volume at SNR 10dB. }
		\label{tab:time-memory}
	\end{center}
	\vspace{-0.5cm}
\end{table}

\subsection{Robustness to Noise}

\begin{figure}
    \centering
	\begin{minipage}{.49\textwidth}
	\begin{tikzpicture}     
		\begin{axis}[
			width=0.9\linewidth, 
			grid=major, 
			grid style={dashed,gray!30}, 
			xlabel= SNR (dB),
			ylabel=correlation coefficient (CC),
			xtick={-20,-10,0,10,20,30},
			xmin = -20, xmax = 30,
			ymin = 0, ymax=1,
			axis x line*=bottom,
			axis y line*=left,
			legend style={at={(0.42,1.15)}, legend cell align=left, align=left, draw=none,font=\scriptsize},
			xticklabel style={
				/pgf/number format/fixed,
				/pgf/number format/precision=0,
				/pgf/number format/fixed zerofill
			},
			scaled x ticks=false
			]
			\addplot[color=black,line width=2*\lw, opacity=\a] table [x=SNR, y=icetide, col sep=comma] {images/SNR_exp/CC.csv};
			\addlegendentry{ICE-TIDE}
			\addplot[color=black,line width=\lw, opacity=\a, style=densely dashed] table [x=SNR, y=ETOMO, col sep=comma] {images/SNR_exp/CC.csv};
			\addlegendentry{Etomo}
			\addplot[color=black,line width=\lw, opacity=\a] table [x=SNR, y=AreTomo, col sep=comma] {images/SNR_exp/CC.csv};
			\addlegendentry{AreTomo}
			\addplot[color=black,line width=\lw, opacity=\a,mark=o, mark size=1] table [x=SNR, y=tv_no_deformed, col sep=comma] {images/SNR_exp/CC.csv};
			\addlegendentry{TV w/o def}
			\addplot[color=black,line width=\lw, opacity=\a,mark=+, mark size=2] table [x=SNR, y=FBP_no_deformed, col sep=comma] {images/SNR_exp/CC.csv};
			\addlegendentry{FBP w/o def}
		\end{axis}   
	\end{tikzpicture}	
	\caption{Correlation coefficient of reconstructions of the SHREC \textit{model 0} volume for different SNRs.\label{fig:noise}} 
    \end{minipage}
\end{figure}
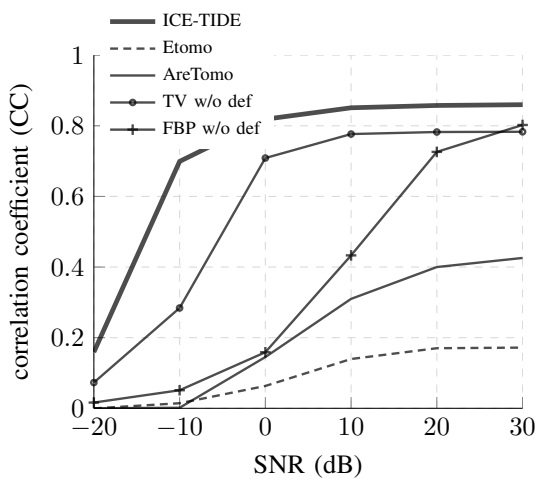

Cryo-ET data have a very low SNR because biological specimens cannot be exposed to high-dose electron beams. Reconstruction algorithms should thus be robust to noise. Here we assess the quality of reconstruction for different levels of noise on the tilt series. 
We report the correlation coefficient in Fig. \ref{fig:noise} experimenting with SNRs ranging from $-20$dB to $30$dB. 
While high SNR data is not practically viable, it gives a sense of the level of detail that ICE-TIDE can theoretically capture if noise could be attenuated by any means. Correlation obtained with ICE-TIDE stabilizes around 0.9 at moderate SNR ($>10$dB), whereas it still increases significantly in pure FBP methods or AreTomo.

\subsection{Training Epochs}
\begin{figure}
	\begin{minipage}{.49\textwidth}	
    \begin{tikzpicture}     
		\begin{axis}[
			width=0.9\linewidth, 
			grid=major, 
			grid style={dashed,gray!30}, 
			xlabel= Epochs,
			ylabel=correlation coefficient (CC),
			xmin = 0, xmax = 1000,
			ymin = 0, ymax=1,
			axis x line*=bottom,
			axis y line*=left,
			legend style={at={(1,0.75)}, legend cell align=left, align=left, draw=none,font=\scriptsize},
			xticklabel style={
				/pgf/number format/fixed,
				/pgf/number format/precision=0,
				/pgf/number format/fixed zerofill
			},
			scaled x ticks=false
			]
			\addplot[color=black,line width=2*\lw, opacity=\a, mark=halfdiamond*, mark size=1] table [x=ep, y=icetide, col sep=comma] {images/volume_epochs/CC_iter.csv};
			\addlegendentry{ICE-TIDE}
			\addplot[color=black,line width=\lw, opacity=\a, mark=otimes*] table [x=ep, y=TV_no_deformed, col sep=comma] {images/volume_epochs/CC_iter.csv};
             \addplot[color=black,line width=\lw, opacity=\a,mark=+, mark size=2] table [x=ep, y=FBP_no_deformed, col sep=comma] {images/volume_epochs/CC_iter.csv};

             \addlegendimage{color=mycolor3,line width=\lw, opacity=\a, mark=otimes*}
             \addlegendentry{TV w/o def}
             \addlegendimage{color=mycolor4,line width=\lw, opacity=\a,mark=+, mark size=2}
             \addlegendentry{FBP w/o def }
		\end{axis}   
	\end{tikzpicture}	
	\caption{ICE-TIDE reconstruction quality on the SHREC \textit{model 0} volume at SNR 10dB over training epochs as measured by the correlation coefficient. \label{fig:epocs}} 
    \end{minipage}
\end{figure}
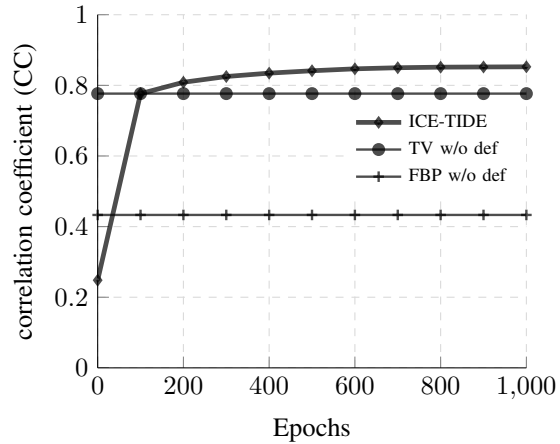

One of the important hyperparameters when estimating the volume with our method  is the number of training iterations (epochs). While increasing the number of epochs generally tends to result in better reconstructions, it proportionally affects the runtime of ICE-TIDE.
In Fig. \ref{fig:epocs}, we report the correlation coefficient of the estimated volumes at intermediate stages compared to the ground truth volume, at intervals of $100$ epochs.
In this experiment, an epoch is equivalent to scanning 1500 pixels taken at random from each projection in the tilt-series. 
We observe that the resolution increases with the number of epochs; \revision{$200$ epochs already yield the correlation better than that of TV without deformations.} Again, this can be attributed to the favorable inductive bias of the implicit neural network which effectively compensates for the missing wedge but also the sparsity of views in the observed wedge. To ensure robustness when comparing with other parameters, we trained the model for a total of $1000$ epochs.

\subsection{Multiple Volume Comparison}
The SHREC 2021 dataset contains 10 synthetic volumes with different distribution of proteins within  "cells", named \textit{model 0} to \textit{model 9}. In order to show that the findings described in previous paragraphs generalize to other volumes, we simulate deformations and projections and compare the reconstructions for all 10 volumes in Fig. \ref{fig:shre-res-05}.
We emphasize that similar studies on a large collection of different volumes are rare in cryo-ET, where the availability of high-quality ground truth data is very limited.
We can see that the results obtained by ICE-TIDE are consistently accurate across all the volumes.

\def\a{0.4}
\begin{figure*}
\centering
\begin{tikzpicture}
  \begin{axis}[
  width=0.8\linewidth,
  height=.3\linewidth,
     ybar,
    bar width=0.20cm,
    axis lines*=left,
    clip=false,
    ylabel={correlation coefficient (CC)},
     xtick=data,
    symbolic x coords={model 0, model 1, model 2, model 3,model 4, model 5, model 6, model 7,model 8, model 9},
     x tick label style={rotate=45, anchor=east},
     legend pos=north east,
     legend style={at={(1.17,0.98)}},
     legend cell align={left},
     legend image code/.code={
        \draw [#1] (0cm,-0.1cm) rectangle (0.2cm,0.25cm); },
    ]
    \addplot[line width=0pt, color=white, fill=mycolor3,fill opacity=\a, bar shift=-0.3cm, 
    postaction={
        pattern=north east lines
    }]  table [x=MODEL_NAME, y=FBP, col sep=comma] {images/shrec_comparison/CC.csv};
    \addlegendentry{FBP}
    \addplot[line width=0pt,color=white, fill=mycolor2,fill opacity=\a, bar shift=0.3cm,
    style = {fill=mycolor2, mark=none, postaction={pattern=horizontal lines}},]  table [x=MODEL_NAME, y=AreTomo, col sep=comma] {images/shrec_comparison/CC.csv};
    \addlegendentry{AreTomo}
    \addplot[line width=0pt,color=white, fill=mycolor4,fill opacity=\a, bar shift=-0.1cm,
    postaction={
        pattern=north west lines
    }]  table [x=MODEL_NAME, y=tv_no_deformed, col sep=comma] {images/shrec_comparison/CC.csv};
    \addlegendentry{TV w/o def}
    \addplot[line width=0pt,color=white, fill=mycolor1,fill opacity=\a, bar shift=0.1cm]  table [x=MODEL_NAME, y=icetide, col sep=comma] {images/shrec_comparison/CC.csv};
    \addlegendentry{ICE-TIDE}
  \end{axis}
\end{tikzpicture}
\caption{Correlation coefficient between ground truth and reconstruction of different SHREC volumes at SNR 10dB.}
\label{fig:shre-res-05}
\end{figure*}
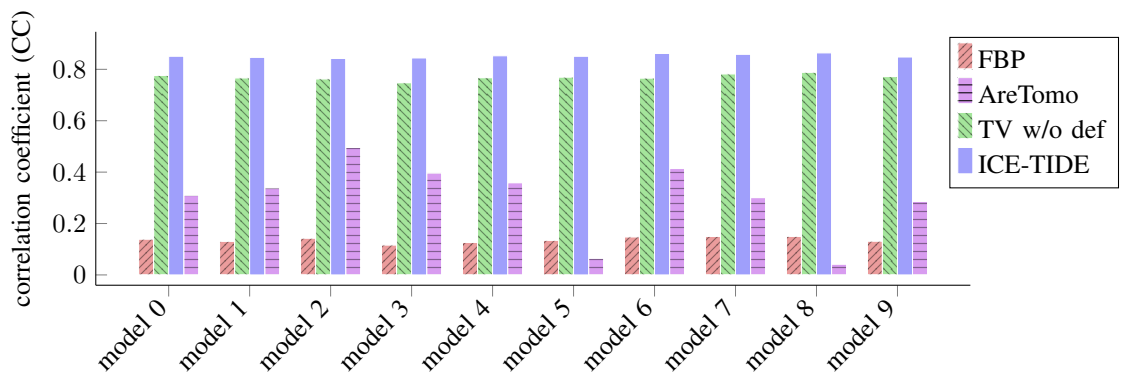

\subsection{Experiments on real data}
\revision{
Finally, we evaluate the performance of ICE-TIDE on real data without ground truth.  We select a tomogram of SARS-CoV-2 virions available in Empiar-11070 \cite{calder2022electron}, as it contains gold fiducials. This enables us to align the tilt series by explicitly utilizing these markers in Etomo, effectively yielding what could correspond to an undeformed tilt series, assuming local deformations are negligible. The tilt-series is binned by a factor of 4 to minimize the effect of the CTF. We use TV reconstructions on the tilt-series, which was aligned using both Etomo and the deformations estimated by \method{} for comparison.
Additionally, we report the estimated volume produced by the implicit neural network. A visual comparison of these reconstruction is shown in Fig. \ref{fig:11070}.

To account for the accumulated electron dose in the tilt series as well as for the increased sample thickness, we estimate a dose correction weight together with the deformations and the volume density. The weights $\{w_m\}$ in the optimization problem \eqref{eq:opt} are chosen to decrease quadratically between 2 for tilt at 0 degrees and 1 for tilt at angles $-60$ and $60$ degrees.

Fig. \ref{fig:11070} displays orthogonal slices of the reconstructed volumes, where the slices are averaged over 15 slices around the central ones to produce visually smoother reconstructions.
We observe that \method{} is capable of effectively aligning the noisy tilt-series, yielding  results comparable to those obtained by aligning the tilt-series using the gold fiducials.

The primary objective of \method{} is to facilitate the alignment of arbitrary datasets that do not contain fiducial markers. This controlled experiment demonstrates that a global deformation closely aligned with the ground truth can be obtained, indicating that \method{} accurately estimates the deformations in real tilt series.

However, it is important to note that, with only minor adaptations to the code tested on simulated data, \method{} is able to achieve decent results on real data that may suffer from different effects not taken into account (CTF, Poisson noise).
To effectively handle real data, it is crucial to consider these physical effects, which requires considerable effort, as discussed in the Discussion section.
}

\def\ss{0.37}
\def\xx{4.74}
\def\yy{-4.12}
\begin{figure*}   
    \centering 
    \begin{subfigure}[t]{0.3\textwidth}
    \begin{tikzpicture}[spy using outlines={circle,orange,magnification=4,size=1.5cm, connect spies}]
        \node[rotate=0, line width=5mm, line width=0.05mm, draw=white] at (0, 0) { \includegraphics[scale=\ss]{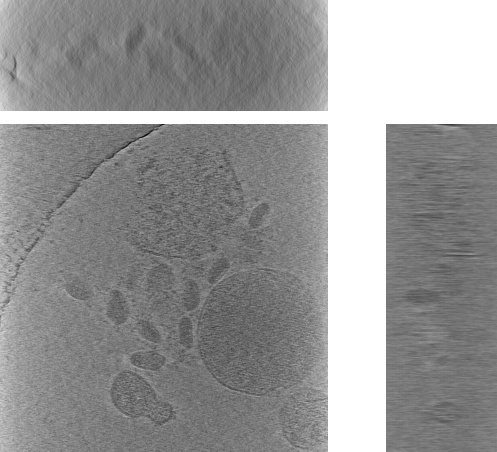}};
    \end{tikzpicture} 
   \caption{TV reconstruction using the aligned tilt-series using fiducial markers.}
    \end{subfigure}\hfill
    \begin{subfigure}[t]{0.3\textwidth}
    \centering
    \begin{tikzpicture}[spy using outlines={circle,orange,magnification=4,size=1.5cm, connect spies}]
        \node[rotate=0, line width=5mm, line width=0.05mm, draw=white] at (0, 0) { \includegraphics[scale=\ss]{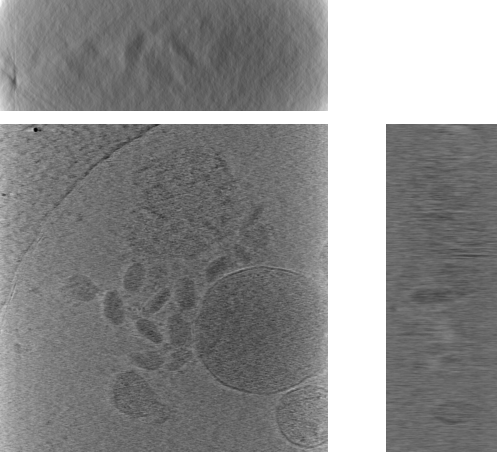}};
    \end{tikzpicture} 
   \caption{TV reconstruction using \method{} estimated deformations.}
   \label{fig:covid_our_tv} 
    \end{subfigure}\hfill
    \begin{subfigure}[t]{0.3\textwidth}
    \centering
    \begin{tikzpicture}[spy using outlines={circle,orange,magnification=4,size=1.5cm, connect spies}]
        \node[rotate=0, line width=5mm, line width=0.05mm, draw=white] at (0, 0) { \includegraphics[scale=\ss]{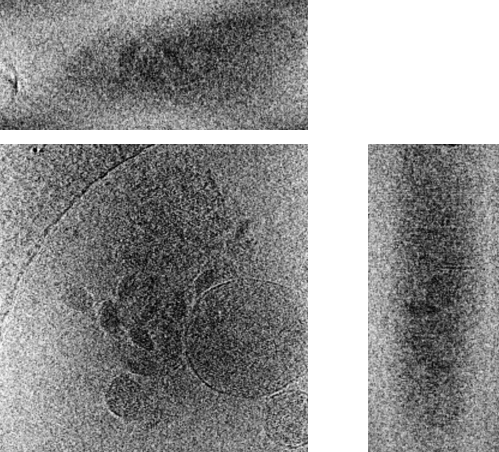}};
    \end{tikzpicture} 
   \caption{\method{} reconstruction.}
   \label{fig:covid_our} 
    \end{subfigure}
    \caption{\revision{Slice reconstructions of Empiar-11070 dataset. Contrast has been stretched
for better visualization.}}
    \label{fig:11070} 
\end{figure*}

\section{Discussion and limitations}

The success of ICE-TIDE is based on a careful modeling of the acquisition pipeline in cryo-ET, including the various sample deformations.
It is remarkable that ICE-TIDE yields high-quality reconstructions both on synthetic measurements, where we can assess the reconstruction quality relative to the ground truth, and on real measurements, even though we have for simplicity not modeled the effect of the CTF and we have considered a simple noise model.
Adapting ICE-TIDE to compensate for the effect of the CTF will require some original thinking. A straightforward extension of \eqref{eq:cryo-ET_formulation}, assuming that the CTF has support of size $c\times c$ pixels in real space, would  increase computational cost by a factor of $c^2$---a prohibitive slow down given that $c$ should be taken comparatively large.

\revision{The CTF effects are especially important in thick samples, as usually encountered in cryo-ET, due to the large defocus gradient introduced. Moreover, for thick specimens, it is known that deformations are better modeled on the 3D volumes directly \cite{brilot2012beam, fernandez2018cryo, fernandez2019consideration}, contrary to what is done in ICE-TIDE. This more realistic modeling yields also a more challenging computational problem that can only be tackled while using additional physical prior, e.g. modeling relation between deformations at different tilts. Taking into account such 3D effects as well as the CTF is crucial to reconstruct high resolution volumes.}

For low-electron regime acquisitions, noise is better modeled by a random variable dependent on signal intensity such as Poisson.
However, this modeling mismatch does not seem to significantly affect the quality of the reconstructions, as observed in our experiments.
Nonetheless, more accurate noise models can be implemented simply by changing the data fidelity term \eqref{eq:data}, see \cite{figueiredo2010restoration}. 
Taking into account low-electron regimes opens up the possibility of extending the tilt series alignment and reconstruction process to directly read dose-fractionated data, i.e. the raw movie frames obtained at each viewing angle within a short exposure time, which are currently aligned and averaged at a separate stage beforehand. Note that this would also increase the required computing resources proportionally to the number of fractions.

Although we do not explicitly impose prior knowledge constraints for estimating the missing wedge information, we observe that ICE-TIDE estimates can retrieve this information with some accuracy. 
As already mentioned, this is due to the parameterization of the volume density by an implicit neural network.
It is also possible to infer information from the missing wedge by assuming, for example, that it is equally likely to observe each patch of the volume of interest or any of the rotations of that patch \cite{chen2021equivariant}.
This assumption makes sense from a physical viewpoint, as many features in the volume such as proteins are indeed approximately equally likely to appear at any orientation.
Such a prior could be incorporated as an additional regularization term in \eqref{eq:opt}. This is related to ideas proposed in IsoNet \cite{liu2022isotropic} and DeepDeWedge \cite{wiedemann_deep_2023}.

Finally, we emphasize that ICE-TIDE is implemented so that it can be used on a desktop computer with a GPU. With some engineering, the current code could probably be adapted to obtain a significant gain in computation time, at the cost of an increased GPU-memory consumption.

\section{Conclusion}
We presented ICE-TIDE, an original method to address the problem of cryo-ET reconstruction in the presence of physical deformations affecting the noisy projections. ICE-TIDE leverages an implicit neural network architecture that is able to learn with remarkable fidelity the underlying 3D volume. Interestingly, our method can not only provide accurate tilt series alignments but also directly yield the tomogram. Furthermore, the network learns to restore information not directly recorded in the tilt series due to the discrete and range-limited nature of the tilting scheme possible in cryo-ET, thus compensating for the missing wedge to considerable extent.

Local and global deformations may have different temporal resolutions. We ignore that which makes our model more expressive and able to recover the specimen imaged with great fidelity. More accurate knowledge of the temporal nature of the deformations can be further incorporated as \textit{a priori} knowledge in our method.

We show that the unknown parameters of the model can be estimated simply by performing a gradient descent on a data fidelity loss. While we focused here on the challenging problem of cryo-ET, the only requirement is the precise knowledge of the forward operator. Consequently, more general inverse problems could be addressed with this formalism.

\bibliographystyle{unsrt}
\bibliography{refs}

\begin{thebibliography}{10}

\bibitem{shkolnisky2012viewing}
Yoel Shkolnisky and Amit Singer.
\newblock Viewing direction estimation in cryo-em using synchronization.
\newblock {\em SIAM journal on imaging sciences}, 5(3):1088--1110, 2012.

\bibitem{wan2016cryo}
W~Wan and John~AG Briggs.
\newblock Cryo-electron tomography and subtomogram averaging.
\newblock {\em Methods in enzymology}, 579:329--367, 2016.

\bibitem{liu2022isotropic}
Yun-Tao Liu, Heng Zhang, Hui Wang, Chang-Lu Tao, Guo-Qiang Bi, and Z~Hong Zhou.
\newblock Isotropic reconstruction for electron tomography with deep learning.
\newblock {\em Nature communications}, 13(1):6482, 2022.

\bibitem{wiedemann_deep_2023}
Simon Wiedemann and Reinhard Heckel.
\newblock A {Deep} {Learning} {Method} for {Simultaneous} {Denoising} and
  {Missing} {Wedge} {Reconstruction} in {Cryogenic} {Electron} {Tomography},
  November 2023.
\newblock arXiv:2311.05539 [cs].

\bibitem{campbell2012movies}
Melody~G Campbell, Anchi Cheng, Axel~F Brilot, Arne Moeller, Dmitry Lyumkis,
  David Veesler, Junhua Pan, Stephen~C Harrison, Clinton~S Potter, Bridget
  Carragher, et~al.
\newblock Movies of ice-embedded particles enhance resolution in electron
  cryo-microscopy.
\newblock {\em Structure}, 20(11):1823--1828, 2012.

\bibitem{brilot2012beam}
Axel~F Brilot, James~Z Chen, Anchi Cheng, Junhua Pan, Stephen~C Harrison,
  Clinton~S Potter, Bridget Carragher, Richard Henderson, and Nikolaus
  Grigorieff.
\newblock Beam-induced motion of vitrified specimen on holey carbon film.
\newblock {\em Journal of structural biology}, 177(3):630--637, 2012.

\bibitem{zheng2022aretomo}
Shawn Zheng, Georg Wolff, Garrett Greenan, Zhen Chen, Frank~GA Faas, Montserrat
  B{\'a}rcena, Abraham~J Koster, Yifan Cheng, and David~A Agard.
\newblock Aretomo: An integrated software package for automated marker-free,
  motion-corrected cryo-electron tomographic alignment and reconstruction.
\newblock {\em Journal of Structural Biology: X}, 6:100068, 2022.

\bibitem{russo_charge_2018}
Christopher~J. Russo and Richard Henderson.
\newblock Charge accumulation in electron cryomicroscopy.
\newblock {\em Ultramicroscopy}, 187:43--49, April 2018.

\bibitem{andersen1984simultaneous}
Anders~H Andersen and Avinash~C Kak.
\newblock Simultaneous algebraic reconstruction technique (sart): a superior
  implementation of the art algorithm.
\newblock {\em Ultrasonic imaging}, 6(1):81--94, 1984.

\bibitem{gilbert1972iterative}
Peter Gilbert.
\newblock Iterative methods for the three-dimensional reconstruction of an
  object from projections.
\newblock {\em Journal of theoretical biology}, 36(1):105--117, 1972.

\bibitem{feldkamp1984practical}
Lee~A Feldkamp, Lloyd~C Davis, and James~W Kress.
\newblock Practical cone-beam algorithm.
\newblock {\em Josa a}, 1(6):612--619, 1984.

\bibitem{kremer1996computer}
James~R Kremer, David~N Mastronarde, and J~Richard McIntosh.
\newblock Computer visualization of three-dimensional image data using imod.
\newblock {\em Journal of structural biology}, 116(1):71--76, 1996.

\bibitem{chen_complete_2019}
Muyuan Chen, James~M. Bell, Xiaodong Shi, Stella~Y. Sun, Zhao Wang, and
  Steven~J. Ludtke.
\newblock A complete data processing workflow for cryo-{ET} and subtomogram
  averaging.
\newblock {\em Nature Methods}, 16(11):1161--1168, November 2019.
\newblock Publisher: Springer US.

\bibitem{de2022scipiontomo}
J~Jim{\'e}nez de~la Morena, Pablo Conesa, Yunior~Cesar Fonseca, Federico~Pedro
  de~Isidro-G{\'o}mez, David Herreros, E~Fern{\'a}ndez-Gim{\'e}nez, David
  Strelak, Emmanuel Moebel, Tim-Oliver Buchholz, Florian Jug, et~al.
\newblock Scipiontomo: Towards cryo-electron tomography software integration,
  reproducibility, and validation.
\newblock {\em Journal of structural biology}, 214(3):107872, 2022.

\bibitem{pyle_current_2021}
Euan Pyle and Giulia Zanetti.
\newblock Current data processing strategies for cryo-electron tomography and
  subtomogram averaging.
\newblock {\em Biochemical Journal}, 478(10):1827--1845, May 2021.

\bibitem{gupta2022differentiable}
Sidharth Gupta, Konik Kothari, Valentin Debarnot, and Ivan Dokmani{\'c}.
\newblock Differentiable uncalibrated imaging.
\newblock {\em IEEE Transactions on Computational Imaging}, 2024.

\bibitem{debarnot2022joint}
Valentin Debarnot, Sidharth Gupta, Konik Kothari, and Ivan Dokmani{\'c}.
\newblock Joint cryo-et alignment and reconstruction with neural deformation
  fields.
\newblock In {\em ICASSP 2023-2023 IEEE International Conference on Acoustics,
  Speech and Signal Processing (ICASSP)}, pages 1--5. IEEE, 2023.

\bibitem{punjani2017cryosparc}
Ali Punjani, John~L Rubinstein, David~J Fleet, and Marcus~A Brubaker.
\newblock cryosparc: algorithms for rapid unsupervised cryo-em structure
  determination.
\newblock {\em Nature methods}, 14(3):290--296, 2017.

\bibitem{li2013electron}
Xueming Li, Paul Mooney, Shawn Zheng, Christopher~R Booth, Michael~B Braunfeld,
  Sander Gubbens, David~A Agard, and Yifan Cheng.
\newblock Electron counting and beam-induced motion correction enable
  near-atomic-resolution single-particle cryo-em.
\newblock {\em Nature methods}, 10(6):584--590, 2013.

\bibitem{Zheng2017}
Shawn~Q Zheng, Eugene Palovcak, Jean-Paul Armache, Kliment~A Verba, Yifan
  Cheng, and David~A Agard.
\newblock {MotionCor2}: anisotropic correction of beam-induced motion for
  improved cryo-electron microscopy.
\newblock {\em Nature Methods}, 14(4):331--332, April 2017.
\newblock Publisher: Nature Publishing Group, a division of Macmillan
  Publishers Limited. All Rights Reserved.

\bibitem{Grant2015}
Timothy Grant and Nikolaus Grigorieff.
\newblock Measuring the optimal exposure for single particle cryo-{EM} using a
  2.6 Å reconstruction of rotavirus {VP6}.
\newblock {\em eLife}, 4, May 2015.

\bibitem{rohou2015ctffind4}
Alexis Rohou and Nikolaus Grigorieff.
\newblock Ctffind4: Fast and accurate defocus estimation from electron
  micrographs.
\newblock {\em Journal of structural biology}, 192(2):216--221, 2015.

\bibitem{tegunov2019real}
Dimitry Tegunov and Patrick Cramer.
\newblock Real-time cryo-electron microscopy data preprocessing with warp.
\newblock {\em Nature methods}, 16(11):1146--1152, 2019.

\bibitem{mastronarde2017automated}
David~N Mastronarde and Susannah~R Held.
\newblock Automated tilt series alignment and tomographic reconstruction in
  imod.
\newblock {\em Journal of structural biology}, 197(2):102--113, 2017.

\bibitem{fernandez_tomoalign_2021}
Jose-Jesus Fernandez and Sam Li.
\newblock {TomoAlign}: {A} novel approach to correcting sample motion and {3D}
  {CTF} in {CryoET}.
\newblock {\em Journal of Structural Biology}, 213(4):107778, December 2021.

\bibitem{bogensperger2022joint}
Lea Bogensperger, Erich Kobler, Dominique Pernitsch, Petra Kotzbeck, Thomas~R
  Pieber, Thomas Pock, and Dagmar Kolb.
\newblock A joint alignment and reconstruction algorithm for electron
  tomography to visualize in-depth cell-to-cell interactions.
\newblock {\em Histochemistry and Cell Biology}, 157(6):685--696, 2022.

\bibitem{lehtinen2018noise2noise}
Jaakko Lehtinen, Jacob Munkberg, Jon Hasselgren, Samuli Laine, Tero Karras,
  Miika Aittala, and Timo Aila.
\newblock Noise2noise: Learning image restoration without clean data.
\newblock {\em arXiv preprint arXiv:1803.04189}, 2018.

\bibitem{buchholz2019cryo}
Tim-Oliver Buchholz, Mareike Jordan, Gaia Pigino, and Florian Jug.
\newblock Cryo-care: content-aware image restoration for cryo-transmission
  electron microscopy data.
\newblock In {\em 2019 IEEE 16th International Symposium on Biomedical Imaging
  (ISBI 2019)}, pages 502--506. IEEE, 2019.

\bibitem{penczek_image_2010}
Pawel~A. Penczek.
\newblock Image {Restoration} in {Cryo}-{Electron} {Microscopy}.
\newblock In {\em Methods in {Enzymology}}, volume 482, pages 35--72. Elsevier,
  2010.

\bibitem{verbeke_self_2024}
Eric~J. Verbeke, Marc~Aurèle Gilles, Tamir Bendory, and Amit Singer.
\newblock Self {Fourier} shell correlation: properties and application to
  cryo-{ET}.
\newblock {\em Communications Biology}, 7(1):1--9, January 2024.
\newblock Number: 1 Publisher: Nature Publishing Group.

\bibitem{mildenhall2021nerf}
Ben Mildenhall, Pratul~P Srinivasan, Matthew Tancik, Jonathan~T Barron, Ravi
  Ramamoorthi, and Ren Ng.
\newblock Nerf: Representing scenes as neural radiance fields for view
  synthesis.
\newblock {\em Communications of the ACM}, 65(1):99--106, 2021.

\bibitem{khorashadizadeh2022funknn}
AmirEhsan Khorashadizadeh, Anadi Chaman, Valentin Debarnot, and Ivan
  Dokmani{\'c}.
\newblock Funknn: Neural interpolation for functional generation.
\newblock In {\em The Eleventh International Conference on Learning
  Representations}, 2022.

\bibitem{vlavsic2022implicit}
Tin Vla{\v{s}}i{\'c}, Hieu Nguyen, AmirEhsan Khorashadizadeh, and Ivan
  Dokmani{\'c}.
\newblock Implicit neural representation for mesh-free inverse obstacle
  scattering.
\newblock In {\em 2022 56th Asilomar Conference on Signals, Systems, and
  Computers}, pages 947--952. IEEE, 2022.

\bibitem{shi2023harpa}
Cheng Shi, Maarten~V de~Hoop, and Ivan Dokmani{\'c}.
\newblock Harpa: High-rate phase association with travel time neural fields.
\newblock {\em arXiv preprint arXiv:2307.07572}, 2023.

\bibitem{sun2021coil}
Yu~Sun, Jiaming Liu, Mingyang Xie, Brendt Wohlberg, and Ulugbek~S Kamilov.
\newblock Coil: Coordinate-based internal learning for tomographic imaging.
\newblock {\em IEEE Transactions on Computational Imaging}, 7:1400--1412, 2021.

\bibitem{muller2022instant}
Thomas M{\"u}ller, Alex Evans, Christoph Schied, and Alexander Keller.
\newblock Instant neural graphics primitives with a multiresolution hash
  encoding.
\newblock {\em ACM Transactions on Graphics (ToG)}, 41(4):1--15, 2022.

\bibitem{kniesel2022clean}
Hannah Kniesel, Timo Ropinski, Tim Bergner, Kavitha~Shaga Devan, Clarissa Read,
  Paul Walther, Tobias Ritschel, and Pedro Hermosilla.
\newblock Clean implicit 3d structure from noisy 2d stem images.
\newblock In {\em Proceedings of the IEEE/CVF Conference on Computer Vision and
  Pattern Recognition}, pages 20762--20772, 2022.

\bibitem{park2021nerfies}
Keunhong Park, Utkarsh Sinha, Jonathan~T Barron, Sofien Bouaziz, Dan~B Goldman,
  Steven~M Seitz, and Ricardo Martin-Brualla.
\newblock Nerfies: Deformable neural radiance fields.
\newblock In {\em Proceedings of the IEEE/CVF International Conference on
  Computer Vision}, pages 5865--5874, 2021.

\bibitem{mastronarde2007fiducial}
David~N Mastronarde.
\newblock Fiducial marker and hybrid alignment methods for single-and
  double-axis tomography.
\newblock In {\em Electron tomography}, pages 163--185. Springer, 2007.

\bibitem{paszke2017automatic}
Adam Paszke, Sam Gross, Soumith Chintala, Gregory Chanan, Edward Yang, Zachary
  DeVito, Zeming Lin, Alban Desmaison, Luca Antiga, and Adam Lerer.
\newblock Automatic differentiation in pytorch.
\newblock 2017.

\bibitem{kim2022zero}
Chaewon Kim, Jaeho Lee, and Jinwoo Shin.
\newblock Zero-shot blind image denoising via implicit neural representations.
\newblock {\em arXiv preprint arXiv:2204.02405}, 2022.

\bibitem{chakrabarty2019spectral}
Prithvijit Chakrabarty.
\newblock The spectral bias of the deep image prior.
\newblock In {\em Bayesian Deep Learning Workshop and Advances in Neural
  Information Processing Systems (NeurIPS)}, 2019.

\bibitem{tachella2021neural}
Juli{\'a}n Tachella, Junqi Tang, and Mike Davies.
\newblock The neural tangent link between cnn denoisers and non-local filters.
\newblock In {\em Proceedings of the IEEE/CVF Conference on Computer Vision and
  Pattern Recognition}, pages 8618--8627, 2021.

\bibitem{shrec2021}
Ilja Gubins, Marten~L. Chaillet, Gijs van~der Schot, M.~Cristina Trueba,
  Remco~C. Veltkamp, Friedrich Förster, Xiao Wang, Daisuke Kihara, Emmanuel
  Moebel, Nguyen~P. Nguyen, Tommi White, Filiz Bunyak, Giorgos Papoulias,
  Stavros Gerolymatos, Evangelia~I. Zacharaki, Konstantinos Moustakas, Xiangrui
  Zeng, Sinuo Liu, Min Xu, Yaoyu Wang, Cheng Chen, Xuefeng Cui, and Fa~Zhang.
\newblock {SHREC 2021: Classification in Cryo-electron Tomograms}.
\newblock In Silvia Biasotti, Roberto~M. Dyke, Yukun Lai, Paul~L. Rosin, and
  Remco~C. Veltkamp, editors, {\em Eurographics Workshop on 3D Object
  Retrieval}. The Eurographics Association, 2021.

\bibitem{Harauz1986}
George Harauz and Marin van Heel.
\newblock Exact filters for general geometry three dimensional reconstruction.
\newblock {\em Optik}, 78(4):146--156, 1986.
\newblock ISBN: 0030-4026.

\bibitem{chambolle2011first}
Antonin Chambolle and Thomas Pock.
\newblock A first-order primal-dual algorithm for convex problems with
  applications to imaging.
\newblock {\em Journal of mathematical imaging and vision}, 40:120--145, 2011.

\bibitem{gursoy2014tomopy}
Doga G{\"u}rsoy, Francesco De~Carlo, Xianghui Xiao, and Chris Jacobsen.
\newblock Tomopy: a framework for the analysis of synchrotron tomographic data.
\newblock {\em Journal of synchrotron radiation}, 21(5):1188--1193, 2014.

\bibitem{ulyanov2018deep}
Dmitry Ulyanov, Andrea Vedaldi, and Victor Lempitsky.
\newblock Deep image prior.
\newblock In {\em Proceedings of the IEEE conference on computer vision and
  pattern recognition}, pages 9446--9454, 2018.

\bibitem{calder2022electron}
Lesley~J Calder, Thomas Calcraft, Saira Hussain, Ruth Harvey, and Peter~B
  Rosenthal.
\newblock Electron cryotomography of sars-cov-2 virions reveals cylinder-shaped
  particles with a double layer rnp assembly.
\newblock {\em Communications Biology}, 5(1):1210, 2022.

\bibitem{fernandez2018cryo}
Jose-Jesus Fernandez, Sam Li, Tanmay~AM Bharat, and David~A Agard.
\newblock Cryo-tomography tilt-series alignment with consideration of the
  beam-induced sample motion.
\newblock {\em Journal of structural biology}, 202(3):200--209, 2018.

\bibitem{fernandez2019consideration}
Jose-Jesus Fernandez, Sam Li, and David~A Agard.
\newblock Consideration of sample motion in cryo-tomography based on alignment
  residual interpolation.
\newblock {\em Journal of structural biology}, 205(3):1--6, 2019.

\bibitem{figueiredo2010restoration}
M{\'a}rio~AT Figueiredo and Jos{\'e}~M Bioucas-Dias.
\newblock Restoration of poissonian images using alternating direction
  optimization.
\newblock {\em IEEE transactions on Image Processing}, 19(12):3133--3145, 2010.

\bibitem{chen2021equivariant}
Dongdong Chen, Juli{\'a}n Tachella, and Mike~E Davies.
\newblock Equivariant imaging: Learning beyond the range space.
\newblock In {\em Proceedings of the IEEE/CVF International Conference on
  Computer Vision}, pages 4379--4388, 2021.

\bibitem{dietrich_membrane-anchored_2022}
Helge~M. Dietrich, Ricardo~D. Righetto, Anuj Kumar, Wojciech Wietrzynski,
  Raphael Trischler, Sandra~K. Schuller, Jonathan Wagner, Fabian~M. Schwarz,
  Benjamin~D. Engel, Volker Müller, and Jan~M. Schuller.
\newblock Membrane-anchored {HDCR} nanowires drive hydrogen-powered {CO2}
  fixation.
\newblock {\em Nature}, pages 1--8, July 2022.
\newblock Publisher: Nature Publishing Group.

\end{thebibliography}

\appendices

\subsection{Experiments on crowded data without fiducial}

We evaluate the performance of ICE-TIDE on real data without ground truth nor gold particles.  We use a tomogram of \textit{T. kivui}, an anaerobic bacterium that can efficiently perform carbon fixation \cite{dietrich_membrane-anchored_2022}. The raw data set from this study is available at the Electron Microscopy Public Image Archive (EMPIAR-11058). To account for different electron doses in the tilt series, we estimate the dose correction weight together with the deformations and the volume density. The weights $\{w_m\}$ in the optimization problem \eqref{eq:opt} are chosen to decrease quadratically between 2 for tilt at 0 degrees and 1 for tilt at angles $-60$ and $60$ degrees.
\revision{This dataset is particularly challenging for objective evaluations due to the absence of gold fiducials.}
Since the ground truth reconstruction is not known for the experimentally acquired tilt series, we compare the reconstructions from our model with those obtained using a combination of several software packages. The raw tilt series has been  aligned by patch tracking with eTomo, then a first tomogram has been obtained using weighted backprojection in IMOD and then cryo-CARE has been used to denoise using tilt series pairs reconstructed from the odd/even raw frames. The final reconstruction can be found in the Electron Microscopy Data Bank (EMD-15056).
Fig. \ref{fig:tkviui} shows the orthogonal slices of the reconstructed volumes where the slices are averaged over 40 slices around the central ones.
We observe that out of the box ICE-TIDE achieves  a comparable reconstruction quality to that obtained by a trained expert who spends time manually combining several software packages and tuning parameters.
This shows that ICE-TIDE is indeed capable of performing the deformation estimation, denoising and reconstruction steps simultaneously, and even restored the missing wedge and sparse angular sampling to some extent. 
\revision{We emphasize that with minimal adaptation from the code tested on simulated dataset, \method{} is able to reach encouraging results.}
In order to deal seriously with real data, it is necessary to take into account various physical effects (CTF, noise, etc.), which requires considerable work, see the Discussion section.

\def\ss{0.52}
\def\xx{4.74}
\def\yy{-4.12}
\begin{figure}   
    \centering 
    \begin{subfigure}[t]{0.49\textwidth}
    \centering 
    \begin{tikzpicture}[spy using outlines={circle,orange,magnification=4,size=1.5cm, connect spies}]
        \node[rotate=0, line width=5mm, line width=0.05mm, draw=white] at (0, 0) { \includegraphics[scale=\ss]{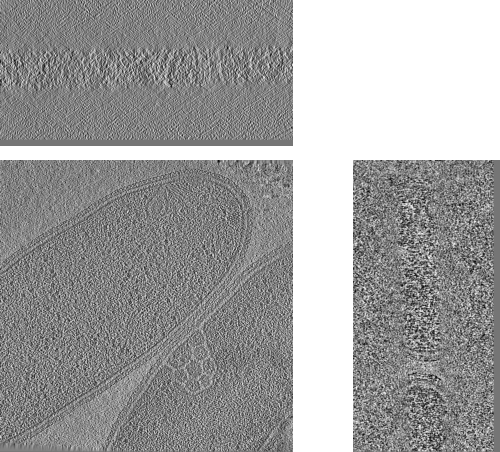}};
    \end{tikzpicture} 
   \caption{Reconstruction using a sequence of specialized software packages.}
    \end{subfigure}\\
    \centering 
    \begin{subfigure}[t]{0.49\textwidth}
    \centering
    \begin{tikzpicture}[spy using outlines={circle,orange,magnification=4,size=1.5cm, connect spies}]
        \node[rotate=0, line width=5mm, line width=0.05mm, draw=white] at (0, 0) { \includegraphics[scale=\ss]{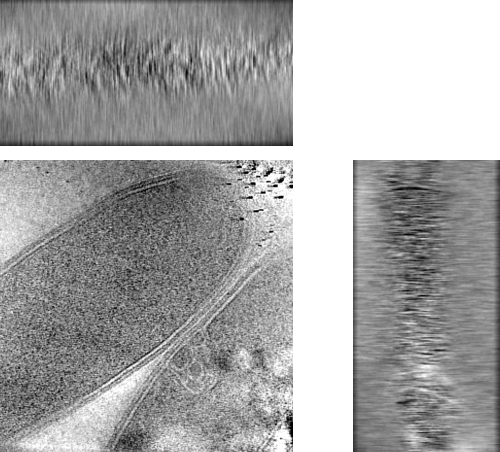}};
    \end{tikzpicture} 
   \caption{ICE-TIDE}
   \label{fig:tkviui_our} 
    \end{subfigure}
    \caption{Slice of a \textit{T. kivui} tomogram reconstructed using combination of softwares and ICE-TIDE. Contrast has been stretched for better visualization.}
    \label{fig:tkviui} 
\end{figure}

\end{document}